\documentclass[review,11pt]{elsarticle}

\usepackage{amsmath}
\usepackage{amsfonts}
\usepackage[noend]{algpseudocode}
\usepackage{amssymb}
\usepackage[thmmarks,amsmath]{ntheorem}
\usepackage{mathrsfs}
\usepackage{fancyhdr}
\usepackage{algorithm}
\usepackage[english]{babel}
\usepackage{t1enc}
\usepackage[utf8]{inputenc} 
\usepackage{ccaption}
\usepackage{psfrag}
\usepackage{color}
\usepackage{epic}
\usepackage{eepic}
\usepackage{rotating}
\usepackage{enumerate}
\usepackage{array}
\usepackage[titletoc,title]{appendix}
\usepackage{pslatex}
\usepackage{ulem}
\usepackage{comment}
\usepackage{multirow}
\usepackage{graphicx}
\usepackage{chronology}
\usepackage{scrextend}
\usepackage[caption=false,font=footnotesize]{subfig}
\usepackage[margin=3cm]{geometry}

\usepackage{lineno,hyperref}
\modulolinenumbers[5]

\journal{Expert Systems with Applications}
\date{}

\biboptions{authoryear}

\usepackage{pdfpages}
\begin{document}

\begin{frontmatter}
\setcounter{page}{1}
\title{Social Network Analytics for Churn Prediction in Telco: Model Building, Evaluation and Network Architecture}

\cortext[cor1]{Corresponding author}
\author[ru]{Mar\'{i}a \'{O}skarsd\'{o}ttir}\corref{cor1}
\ead{mariaoskars@ru.is}
\author[west]{Cristi\'{a}n Bravo}
\ead{cbravoro@uwo.ca}
\author[vub]{Wouter Verbeke}
\ead{wouter.verbeke@vub.ac.be}
\author[arg]{Carlos Sarraute}
\ead{charles@grandata.com}
\author[feb,sth]{Bart Baesens}
\ead{bart.baesens@kuleuven.be}
\author[feb]{Jan Vanthienen}
\ead{jan.vanthienen@kuleuven.be}
\address[ru]{Department of Computer Science, Reykjavik University, Reykjavik, Iceland.}
\address[west]{Department of Statistical and Actuarial Sciences, The University of Western Ontario, London, Canada. }
\address[feb]{Department of Decision Sciences and Information Management, KU Leuven, Leuven, Belgium.}
\address[sth]{Department of Decision Analytics and Risk, University of Southampton, Southampton, UK.}
\address[vub]{Faculty of Economic and Social Sciences and Solvay Business School, Vrije Universiteit Brussel, Brussels, Belgium.}
\address[arg]{Grandata Labs, Buenos Aires, Argentina.}

\begin{abstract}
Social network analytics methods are being used in the telecommunication industry to predict customer churn with great success.
In particular it has been shown that relational learners adapted to this specific problem enhance the performance of predictive models.
In the current study we benchmark different strategies for constructing a relational learner by applying them to a total of eight distinct call-detail record datasets, originating from telecommunication organizations across the world.
We statistically evaluate the effect of relational classifiers and collective inference methods on the predictive power of relational learners, as well as the performance of models where relational learners are combined with traditional methods of predicting customer churn in the telecommunication industry.
Finally we investigate the effect of network construction on model performance; our findings imply that the definition of edges and weights in the network does have an impact on the results of the predictive models.
As a result of the study, the best configuration is a non-relational learner enriched with network variables, without collective inference, using binary weights and undirected networks.
In addition, we provide guidelines on how to apply social networks analytics for churn prediction in the telecommunication industry in an optimal way, ranging from network architecture to model building and evaluation.
 \end{abstract}


\end{frontmatter}


\section{Introduction}\label{secIntroduction}

Customer churn prediction in telecommunication companies (telcos) has become an increasingly popular research topic in the literature in recent years \citep{de2011empirical,modani2013cdr, verbeke2012new}. 
The competitive landscape of these companies, in which customers have many providers to choose from and can easily switch providers should they become unhappy, creates a fierce environment that requires a high level of sophistication to thrive.
Apart from that, studies have shown that customer attrition can be much more expensive for companies than customer retention is \citep{athanassopoulos2000customer,berson2002building}.
A widely used strategy is therefore to identify customers with the highest propensity to churn, and offer them incentives to persuade them to stay.  
Long-term customers are also more profitable for the company, since they are more likely to buy additional products and spread the word of their satisfaction, thus indirectly attracting more customers \citep{ganesh2000understanding}.
Finally, telcos gather an abundance of data about their customers, such as demographics, financials, usage behavior and call records which presents the opportunity to make these data actionable by using analytics techniques.

\begin{figure}
\centering
 \hspace*{\fill}%
\subfloat{\includegraphics[scale=0.15]{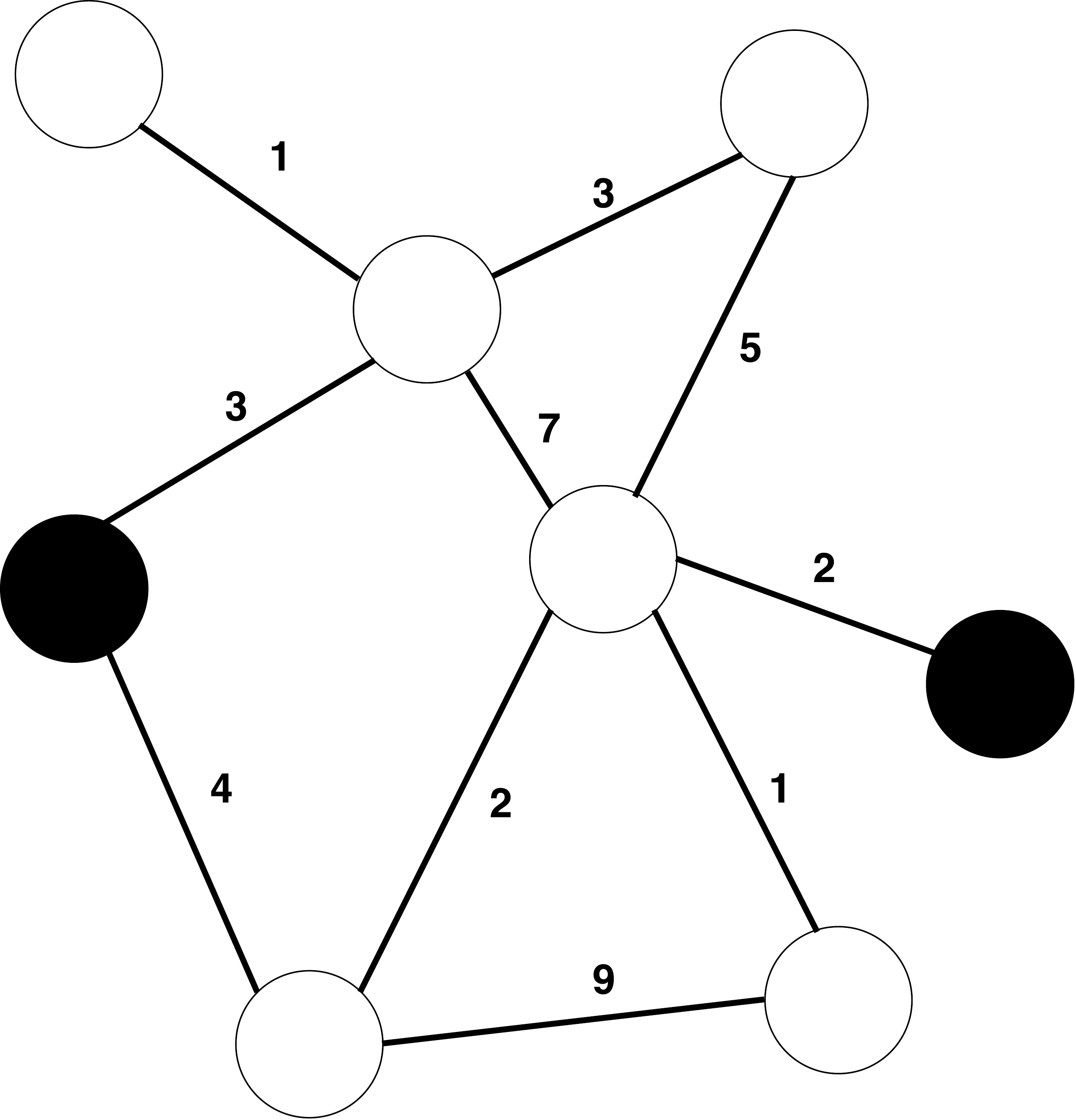}} \hfill
\subfloat{\includegraphics[scale=0.15]{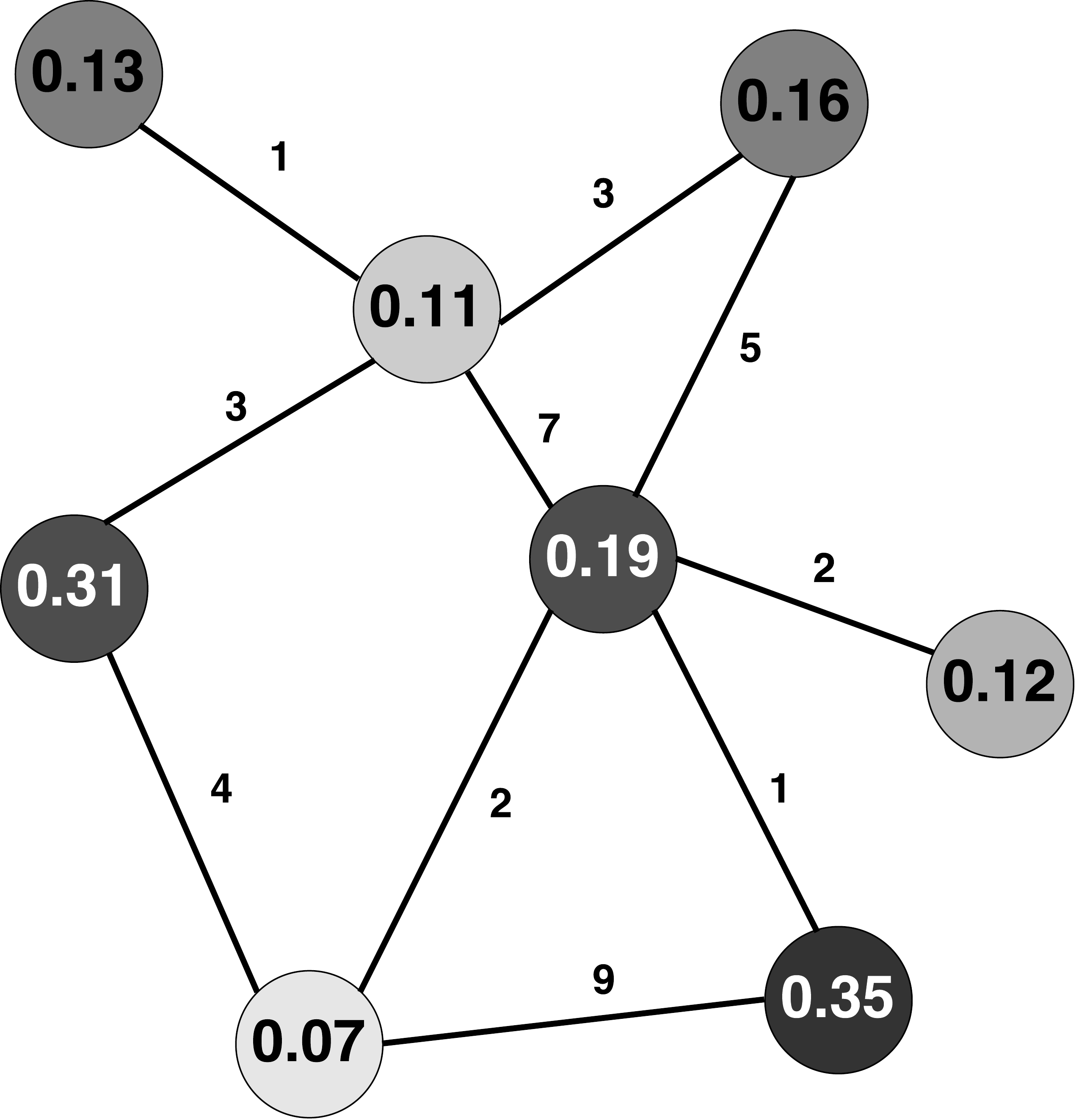}}
 \hspace*{\fill}%
\caption{\label{F:introRL} The figures show an example of an application of a relational learner.  The figure on the left displays a graph with eight customers, of which two have churned (black) and six have not churned (white).  The figure on the right shows the same network after the RL has been applied. Each customer now has a score or probability of churning.}
\end{figure}

In classical customer churn prediction (CCP) modeling, a binary classifier is applied to available customer data at the company to build a predictive model which assigns each customer a score representing their propensity of churning.
Social network analytics (SNA) has become a substantial addition to this field, as studies show that, when the customer datasets contain network features in addition to customer attributes, the performance of CCP models is enhanced \citep{backiel2014mining,kusuma2013combining,richter2010predicting}. 
The network features are extracted from call networks and encapsulate both calling behavior and interactions between customers.
As such they carry valuable information that can be used to generate more accurate CCP models.
In addition, a group of methods called relational -- or network -- learners have been successfully used to exploit the information flow between connected customers in a call network to predict churn \citep{dasgupta2008social,dierkes2011estimating,verbeke2014social}.
In an exploratory study, \citet{verbeke2014social} adapted and applied the network learning framework and toolkit, NetKit, as proposed by \citet{macskassy2007classification}, to classify customers in a call network.
Relational learners, which are made up of relational classifiers 
and collective inference methods, 
simulate how customers who have already churned affect others and how `churn influence' spreads through the network.
Figure \ref{F:introRL} shows an example of a network before and after application of a relational learner.
The result is a score for each customer that can be used as a churn label or an additional feature in the dataset. 
\cite{verbeke2014social} explored both possibilities and showed that some of the relational learners had great potential when it comes to predicting churn.

When applying SNA, one of the biggest challenges is defining the social network using the available relational data.
For telco, these are usually call-detail records (CDR) that need to be filtered and aggregated in an intelligent way,
for which many possibilities exist.
These include, but are not limited to, whether the network should be uni- or bidirectional, or whether the edges should be represented by binary weights or by weights defined by taking into account the length of the call, or the number of calls between two customers.
Differentiation can be made between calls taking place on certain days or at certain times of the day and also between types of customers (retail vs. corporate).
Some studies filter out connections that are non-reciprocal \citep{dasgupta2008social, haenlein2013social}, while others disregard phone calls that last less than a given threshold, since these possibly represent unintentional phone calls \citep{verbeke2014social}. 
These and many other factors need to be taken into consideration when the network is built.
To the best of our knowledge, very few studies investigate the impact of how the network is defined on the techniques which are applied to them, and the resulting findings \citep{zhu2011role,haenlein2013social}. 
This could be a consequence of the fact that CDR datasets often contain millions and even billions of records which makes pre-processing and transforming them into networks both difficult and time consuming.

\begin{figure}
\centering
\includegraphics[width=13cm]{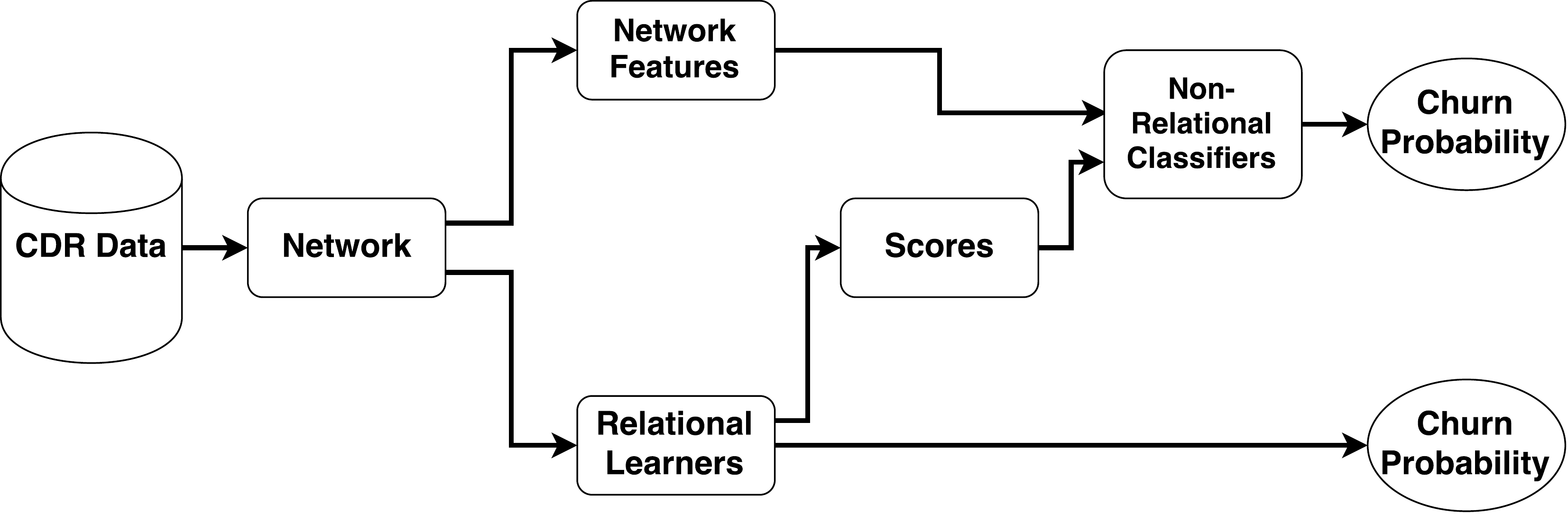}
\caption{\label{F:setup} The model building process}
\end{figure}

\begin{table}
\caption{\label{T:RQ} Research Questions. Performance is measured using four measures: lift at $0.5\%$ and $5\%$, AUC, and EMP as described in subsection \ref{perfM}.}
\centering
\scalebox{0.78}{
\begin{tabular}{|m{1cm}|m{0.5cm}m{17cm}|}
\hline
\multirow{4}{*}{\rotatebox[origin=c]{90}{\parbox[l]{3.5cm}{\centering Effect of   Relational Learners}}}&RQ1&Which relational learners perform statistically different from the rest when predicting customer churn?\\ \cline{2-3}
&RQ2&Do some relational classifiers perform statistically better than the others?\\\cline{2-3}
&RQ3&Is the performance of some collective inference methods statistically better than the others?\\\cline{2-3}
&RQ4&When predicting customer churn, do collective inference methods improve the predictive performance of relational classifiers?\\\hline \hline
\multirow{3}{*}{\rotatebox[origin=c]{90}{\parbox[l]{2.5cm}{\centering Combination of  RL and NRC}}}&RQ5&Which non-relational classifier model performs best when predicting churn?  A model built using network features only, a model with relational learners scores only, or a model which is built with a combination of both?\\\cline{2-3}
&RQ6&Which model type performs better when predicting churn? Relational learners or non-relational classifiers?\\
&&\\\hline
\end{tabular}}
\end{table}
The objective of this study is to explore and evaluate ways of using SNA for churn prediction in telco, from network definition, to model building, and model evaluation and to discuss the implications our results have for academics and practitioners alike.
To achieve this, we compare classification methods that incorporate SNA, and investigate how the network architecture affects the performance of these models.
The study is composed of two main parts.
Firstly, based on \citet{verbeke2014social}, our goal is to rank a selection of relational learners with respect to performance and to see whether combining them with classical non-relational classifiers improves churn prediction in telco.
Although \citet{verbeke2014social} observed differences in performance between the methods, they could not draw conclusions regarding the statistical significance of these differences.
Therefore, we have gathered eight distinct CDR datasets from around the world, which allows us to apply statistical tests for evaluating significance of the results and provide conclusive answers.  
For this part of the analysis we pose six research questions as can be seen in Table \ref{T:RQ}.
The first four questions are targeted towards the relational learners with the aim of ranking them by performance when predicting churn (RQ1), as well as comparing the performance of the two components, relational classifiers (RQ2) and collective inference methods (RQ3), separately.
In addition, because collective inference methods have been shown to improve the performance of relational classifiers in other fields \citep{jensen2004collective}, we investigate whether this is also the case when predicting churn in telco (RQ4). 
Thus, we provide a detailed analysis of the various dimensions of the relational learners in terms of performance. 
Research question RQ5 examines whether combining relational learners with non-relational classifiers improves performance and finally, with research question RQ6 we study how relational learners perform in comparison to non-relational classifiers, and thus, whether they are good enough to be used on their own.
The research questions have practical implications for practitioners in the telecommunications industry who can use them as a guideline on how to optimally apply SNA in churn prediction modelling.
The consequences of the research questions are also scientific, due to the comprehensive analysis of the relational learners using a unique number of CDR datasets.
Figure \ref{F:setup} shows how the churn prediction models were set up, with regards to interaction of relational learners and non-relational classifiers.
For this part of the study, we enforce certain restrictions on the network building and non-relational classifiers because it would be infeasible to test every possible approach.
These decisions are made based on existing literature as detailed later.
Since CDR are the only available source of data, we could not make use of alternative customer attributes, such as socio-demographic data, and marketing and financial information, which are commonly used in the literature \citep{ahn2006customer,huang2012customer,verbeke2012new}.
Again, to the best of our knowledge, no benchmarking study of this scope, using this many CDR datasets, has been conducted.
The second part of this study explores the effect of network construction on model performance.
Using the best performing relational learner on networks defined with increasing complexity using a variation in definition of edges and weights, we study the difference in performance of the resulting classifier.
To the best of our knowledge, this study is the first to explore the impact of network composition on the predictive capability of a SNA-based model.
In summary, our study builds on and extends the previous exploratory study by \citet{verbeke2014social} by exploring the impact of the network architecture or definition on model performance by studying different time frames and representation of edges and weights.
Additionally, we expand the number of data sources involved, allowing us to statistically test and evaluate the results that are found and, as such, to draw conclusive answers to the above-positioned research questions.

We make two main contributions to the field.
Firstly, we provide answers to the six research questions in Table \ref{T:RQ} by empirically evaluating the difference in performance of a set of relational learners and their combination with non-relational classifiers using the eight distinct CDR datasets. 
Thus, we provide a ranking of the relational learners, determine whether relational classifiers perform better when used with collective inference methods, and provide conclusive answers on how to combine relational and non-relational methods to achieve optimum performance.
Secondly, our follow-up study shows that the performance of CCP models does in fact depend on how the network is defined.
Our results imply that the complexity of the phenomenon is not in the complexity of the weights, but in the network structure and the relationship between this structure and the other variables in the dataset.

The rest of this paper is arranged as follows.
In the next section, we discuss the related literature on churn prediction in telco and network construction. 
Next, we describe the methodology of the current research in detail followed by a section about the experimental design that we apply.
Thereafter, we present the results of the experiments together with a discussion about those results.
In section \ref{sec:expost}, we demonstrate how the model is affected by the network, and subsequently we discuss the managerial and academic insights of our results.
The paper concludes with a summary of our main findings and suggestions for future research.

\section{Related Work}
\begin{sidewaystable}
\centering
\caption{\label{T:networkTable} Literature overview of the application of social network analytics for customer churn in telco.}
\scalebox{0.65}{
\footnotesize{
\begin{tabular}{m{3cm}p{8cm}cp{5cm}ccp{5cm}c}
&&&&\multicolumn{4}{c}{Network Details}\\\cline{5-8}
Authors (Year)&Title&$\#$ Datasets&Methods/Application&\# Nodes&Edges&Weights&Duration\\ \hline
\citet{phadke2013prediction}&Prediction of subscriber churn using social network analysis&1&Churn prediction, influence propagation&$5\cdot 10^5$&undirected&duration and number of calls, neighborhood overlap overlap&2 months\\
\citet{richter2010predicting}& Predicting customer churn in mobile networks through analysis of social groups&1&Group first churn prediction&$16\cdot 10^6$&outgoing&&1 month\\
\citet{dasgupta2008social}& Social Ties and their Relevance to Churn in Mobile Telecom Networks&1&Spreading Activation&$2.1\cdot 10^6$&outgoing&call duration&1 month\\
\citet{kim2014improved}& Improved churn prediction in telecommunication industry by analyzng a large network&1&Combination of local features and SPA with logistic regression&$2.4\cdot 10^6$&undirected&duration of calls&2 months\\
\citet{dierkes2011estimating} &Estimating the effect of word of mouth on churn and cross-buying in the mobile phone market with Markov logic networks&1&Logistic regression ,Markov logic with local and network features&$1.2\cdot 10^5$&undirected&number and duration of phonecalls&10 months\\
\citet{zhang2012predicting}&Predicting customer churn through interpersonal influence&1&RL(propagation) with LogReg, DT, NN.&$1\cdot 10^6$&&&7 months\\
\citet{modani2013cdr}&CDR Analysis Based Telco Churn Prediction and Customer Behavior Insights: A Case Study&2&Network featurization, CHAID and LogReg&$12\cdot 10^6$&undirected&number and duration of calls&3 months\\
\citet{kusuma2013combining}&Combining customer attribute and social network mining for prepaid mobile churn prediction &2&Combinations of SPA, network and local variables together with CHAID and LogReg&&undirected&number of calls and texts with decay&1 month\\ 
\citet{verbeke2014social}&Social Network analysis for customer churn predictions&2&Combinations of RL and NRC&$1.2/1.4\cdot 10^6$&undirected&duration of calls&6 months \\
\citet{backiel2014mining}&Mining Telecommunication Networks to Enhance Customer Lifetime Predictions&1&network featurization, LogReg&$1.4\cdot 10^6$&undirected&duration of phonecalls&6 months\\ 
\citet{baras2014effect}&The effect of social affinity and predictive horizon on churn prediction using diffusion modelling&2&RL:SPA, time until churn &&outgoing&Number of calls&2 months\\
\citet{backiel2015combining}&Combining Local and Social Network Classifiers to Improve Churn Prediction&1&Combination of SPA, network variables, non-relational classifers&$1\cdot 10^6$&outgoing&duration of calls&6 months\\ \
\citet{haenlein2013social}&Social interactions in customer churn decisions: The impact of relationship directionality&1&Churn Prediction&$3.5\cdot 10^3$&incoming/outgoing&duration of calls&2 months\\
\citet{rehman2015customer}&Customer churn prediction, segmentation and fraud detection in telecommunication industry&2&Churn prediction&$20\cdot 10^6$&undirected&duration of calls&6/12 months\\ \hline
\citet{nanavati2008analyzing}&Analyzing the structure and evolution of massive telecom graphs&1&&$3\cdot 10^6$&directed&binary, with calls and texts&1 month\\
\citet{miritello2013time}&Time as a limited resource: Communication strategy in mobile phone networks&1&Social behavoir&$20\cdot 10^6$&undirected&duration of calls&11 months\\
\citet{raeder2011predictors}&Predictors of short-term decay of cell phone contacts in a large scale communication network&1&edge prediction&$4.8\cdot 10^6$&directed&decay&2 months\\
\citet{tomar2010social}&Social network analysis of the short message service&\\
\citet{zhu2011role}&Role defining using behavior-based clustering in telecommunication network&1&Behavior Clustering&2&yes&lentgth&1/3Months\\
\end{tabular}
}}
\end{sidewaystable}

As a research domain, CCP is already well established.
This classification problem has been studied intensively in various sectors where maintaining relationships with current customers is considered important. 
It has been applied in the banking sector \citep{ali2014dynamic,glady2009modeling,lariviere2004investigating,van2004customer,xie2009customer}, by insurance companies \citep{guillen2012time,gunther2014modelling}, internet service providers \citep{khan2010applying}, online social networks \citep{ngonmang2012churn}, and in the telecommunication industry \citep{chen2012hierarchical}, which is the case we discuss here.
We refer to \citet{verbeke2012new} for an overview of commonly used classification techniques for CCP in telco and a benchmarking study of those techniques.

Recent publications where social network analysis has been applied to predict customer churn in telco can be seen in the top part of Table \ref{T:networkTable}.
These studies show how the performance of CCP models is enhanced when network effects are taken into account by means of SNA.
The table also lists how edges and weights are defined in each case, together with the number of nodes in the networks and the duration of the CDR data that are used in each study.

As Table \ref{T:networkTable} explains, the most widely used relational learner is the spreading activation method \citep{dasgupta2008social}.
It is used on its own \citep{backiel2015combining,dasgupta2008social} and to produce scores that are then used as variables in non-relational classifiers, such as logistic regression and decision trees \citep{kim2014improved,kusuma2013combining}.
Other works enrich their datasets with network measures for non-relational classifiers, instead of exploiting relational learners \citep{backiel2014mining,benedek2014importance,modani2013cdr,zhang2012predicting}.
\citet{verbeke2014social} compare the performance of multiple relational learners to the performance of non-relational classifiers, with and without network variables.  
The best model is obtained when predictions from both relational learners and non-relational classifiers with network variables were combined.
Additionally, \citet{backiel2015combining} combine these types of models in various ways, with the result that the combined model, a binary classification built using the scores resulting from non-relational classifiers and relational learners (in this case SPA) as variables, gives the best result.
Contrary to the current study, almost none of the discussed papers include a comparison of relational learners in terms of performance or apply their techniques to  multiple CDR datasets. 

For most of the papers in Table \ref{T:networkTable}, SNA relies on a single network which is defined and constructed only once for the situation at hand.
In most cases, length of phone calls between customers are used as weights, but the directionality of the edges varies. 
All studies rely on in-network customers only, since information about customers of other telcos is rarely available.

Only a handful of studies compare the effects of networks defined in different ways or test the difference in results obtained from different networks as we do here.
For example, \cite{haenlein2013social} studies the dynamics of social interactions for customer churn within a directed network.  
He compares customers' propensity to churn using call networks with incoming calls, outgoing calls and both together, producing an undirected network.
According to his study, a customer is more likely to churn if they have been in contact with a person who already churned, but only if the relationship is outgoing, meaning that the customer calls the churned person, and not the other way around.
In addition, \cite{zhu2011role} apply behavior-based clustering to define roles of the customers in their call network using both incoming and outgoing edges.
The people they identify as potential churners make many phone calls but have low betweenness and closeness. In addition, the length of incoming and outgoing calls is higher than for other customers.

So far we have only discussed papers that exploit CDR data for churn prediction in telco, whereas numerous other applications exist. 
An extensive survey by \cite{naboulsi2015large} on the fast-growing field of mobile traffic analysis and how CDR data are increasingly used in data-mining applications classifies the analyses in the existing literature as social, mobility, and network analyses.
Some of the research that falls into the first of these classes, which deals with human dynamics and social interactions, pays special attention to network construction.
The results of these studies might possibly have interesting implications for problem-setting in our context.
The lower part of Table \ref{T:networkTable} contains a brief overview of papers where CDR data have been used to build networks for SNA in telco with various objectives.

\cite{miritello2013time} examine how callers distribute their time across their social network in relation to their network size and intensity of mobile use.
According to their results, people with more connections spend more time communicating than people with fewer connections and the average time spent on each connection increases as the number of connections increases, up to a threshold of 10 to 40 connections.
In addition, independent of the size of their network, people distribute their time unevenly, thus dedicating a small amount of time to many people and a great deal of time to a few people.
In a study about the decay of edges in a call network, \cite{raeder2011predictors} use weighted data to determine the importance of edge weight for the persistence and decay of connecting between people in a call network.
Using machine learning techniques, their results imply that directed edge weight, reciprocated edge weight and recency of connections are important features when predicting edge decay.

\cite{park2012social} propose a relational learner, similar to the spreading activation algorithm, in order to to validate self-reported demographic data of customers. 
In their network construction they use out-degree, number of outgoing calls and duration of outgoing calls and the results imply that the best performing models are those that use networks with out-degree.

Featurizing networks offers the possibility to compare the importance of various network definitions.
This was the case in \cite{sarraute2015inference} who extract features based on different edge and weight definitions, such as incoming and outgoing calls, length and number of calls, and number of text messages in addition to taking time of day and the day of week into account.
In a subsequent PCA analysis, total number of calls, total duration of calls,
and total number of text messages are shown to explain most of the variance.
This indicates that the activity of users is a good candidate to characterize users' social behavior.

\section{Customer Churn Prediction Using Social Networks Analytics}
\subsection{Networks} \label{subsec:networks}
In this particular predictive analytics framework, SNA-based methods and binary classifiers are applied to assign each customer of a telco to one of two classes: churner or non-churner.
The starting point of conducting social network analysis is the network itself \citep{newman2010networks}. 
A network is composed of nodes and edges, which in this case are the customers of a telecommunications operator and the correspondence between them, respectively.
Formally, the node component $\textbf{V}$ of the network consists of a set of nodes $\mathcal{V}=\{v_1,\dots,v_n\}$ and a label vector $\mathcal{L}=\{l_1,\dots,l_n\}$ where each $l_i\in\mathcal{C} =\{c_1,\dots,c_m\}$ is the class label --in this case churner or non-churner-- of node $v_i$.
We denote by $\mathcal{V}^K$ and $\mathcal{V}^U$ the sets of nodes for which class labels are known and unknown, respectively. 
The known labels are used to infer labels for the unknown ones.

The edge component $\textbf{E}$ of the network consists of two parts, edges and weights. 
The edges, $\mathcal{E}$, are a set of the two-subsets of $\mathcal{V}$, where the edge $e_{ij}\in\mathcal{E}$ represents an existing connection from $v_i$ to $v_j$, i.e. if the two customers have shared a phonecall. 
If an edge exists between nodes $i$ and $j$ we say they are connected, and if 
\[
\forall i,j \in {1,\dots, n}: e_{ij}\in\mathcal{E}\iff e_{ji}\in\mathcal{E}
\]
we say that the network is undirected and directed otherwise.
Incoming and outgoing edges of a node $v_i$ are the sets of nodes of the form $e_{ji}$ and $e_{ij}$, respectively

A non-negative weight, $w_{ij}$, can be associated to each edge to denote the strength of the connection. 
There are various ways to define the weights.
They can be binary, indicating whether two nodes are connected or not, or be assigned some other value based on additional information.
In addition, weights can vary in time by conferring more recent connections higher importance than older connections.
To model this in the network, the weights at time $t$, $w_{ij,t}$ can be exponentially weighted in time by 
\begin{equation}
\label{formula}
(w_{ij})_t= e^{-\gamma t}w_{ij,t} 
\end{equation}
where $\gamma$ is the decay constant. 
We obtain the final weights by aggregating all $(w_{ij})_t$ for the whole time period.
Weighted networks have been successfully used in credit card and social security fraud detection \citep{van2015apate,van2014gotcha}.

Finally, the first order neighborhood $\mathcal{N}_i^1$ of a node $v_i\in \mathcal{V}$ is the set of nodes that are connected to $v_i$
\[
\mathcal{N}_i^1=\{v_i\}\cup\{v_j | e_{i,j}\in \mathcal{E}, j=1,\dots,n \}.
\]

\subsection{Enriching Non-Relational Classifiers with Network Features} \label{nrc}
A network can be used to extract network features in a process called featurization \citep{baesens2015fraud}.
Information from the neighborhood of each node, such as the labels of connected nodes and weights of the edges between them, is used to compute numerical attributes to characterize and describe the nodes.
Examples of such features include the number of neighbors a node has (degree), the number of fully connected subgraphs of three nodes (triangles), relational neighbor score, and probabilistic relational neighbor score.
Various features describing the position and connectivity of a node within a network also exist \citep{newman2010networks}.

Link-based features are a specific group of network features, introduced by \citet{lu2003link}. 
These features are constructed using class labels, in addition to nodes and edges.
They define three types of features: mode-link, count-link and binary-link.  
The first one, \textit{mode-link}, is a single feature defined as the most frequently occurring class label amongst the nodes in the neighborhood.
The \textit{count-link} statistic counts the number of times each different class label appears in the neighborhood, and the last statistic, the \textit{binary-link}, documents for each class in $\mathcal{C}$ whether or not it appears as a class label for a node in the neighborhood. 

Featurization has been successfully used to improve the performance of churn prediction models in telco; see Table \ref{T:networkTable}.
Typically, the extracted network features are added to the customer dataset, before binary classifiers, such as decision trees, logistic regression, support vector machines and artificial neural networks are used to train and test CCP models using the enriched dataset.
We limit the number of binary classifiers, since our goal is not to compare them.
Instead, we choose three that have been shown to perform well when predicting churn in telco \citep{verbeke2012new}. 
We choose the classifiers logistic regression (Log) \citep{coussement2008churn,mozer2000predicting}, random forests (RF) \citep{anil2008predicting}, and artificial neural networks (NN) \citep{hung2006applying} because they represent different aspects of the trade-off between complexity and predictive capability \citep{verbraken2014development}.
Popular in the industry, logistic regression is simple to understand, whereas NN and RF are more powerful but harder to interpret since they are black-box models.
In the model building we refer to these classifiers as non-relational classifiers (NRC).
Alternative binary classifiers were implemented as well, for example Adaboost.M1 which has been shown to outperform other methods for CCP in telco \citep{vafeiadis2015comparison}.
However, its performance did not vary significantly from the classifiers mentioned above and it was therefore not included in our experiments.

\subsection{Relational Learning in Social Networks} \label{secRL}
Contrary to featurization for non-relational classifiers, churn probabilities can be inferred from the network directly by exploiting the information flow between the interlinked entities.
Relational learning (RL) for CCP in telco is particularly interesting because it simulates how `churn-influence' spreads through the network and, as a result, how churners might affect non-churners.
In general, relational learners are composed of two parts: relational classifiers and collective inference methods.

\textit{Relational Classifiers} (RC) are the methods which infer class labels for each node in a network based on the weight of links to other nodes and the labels of those nodes.
They perform a single, local operation going from node to node until all have been classified.
The four relational classifiers applied here are the weighted vote relational classifier (WVRN), class distribution relational classifier (CDRN), network-only link-based classifier (NLB), and the spreading activation relational classifier (SPA RC).
A detailed description of each RC is provided in Appendix \ref{App:RL} 
as our goal is not to focus on the methods themselves, but to rank them.

When going through the network in this manner, classifications may not be very stable.
Once the first node has been classified, its label is used to infer the label for the second node, which in turn might change, which could again have an effect on the first node.  
To capture this behavior, collective inference methods are applied. 

\textit{Collective Inference} methods (CI) are procedures which infer class labels for the nodes in a network while taking into account how the inferred labels affect each other.
They decide in which order the nodes are labeled and how a final label is determined.
They have been shown to improve the performance of relational classifiers in genomes and bibliographic networks \citep{jensen2004collective,sen2008collective}.
In general, CIs  perform two operations iteratively until a terminating condition is met.
First, a relational classifier is applied to each node in the network and then the resulting scores are used to update the labels of the nodes.
The methods used here are Gibbs sampling (Gibbs), iterative classification (IC), relaxation labeling (RL), relaxation labeling with simulated annealing (RLSA), and spreading activation collective inference method (SPA CI). 
We refer to Appendix \ref{App:RL} for descriptions and pseudo codes for the CI methods.

\begin{figure}
\centering
\includegraphics[scale=0.6]{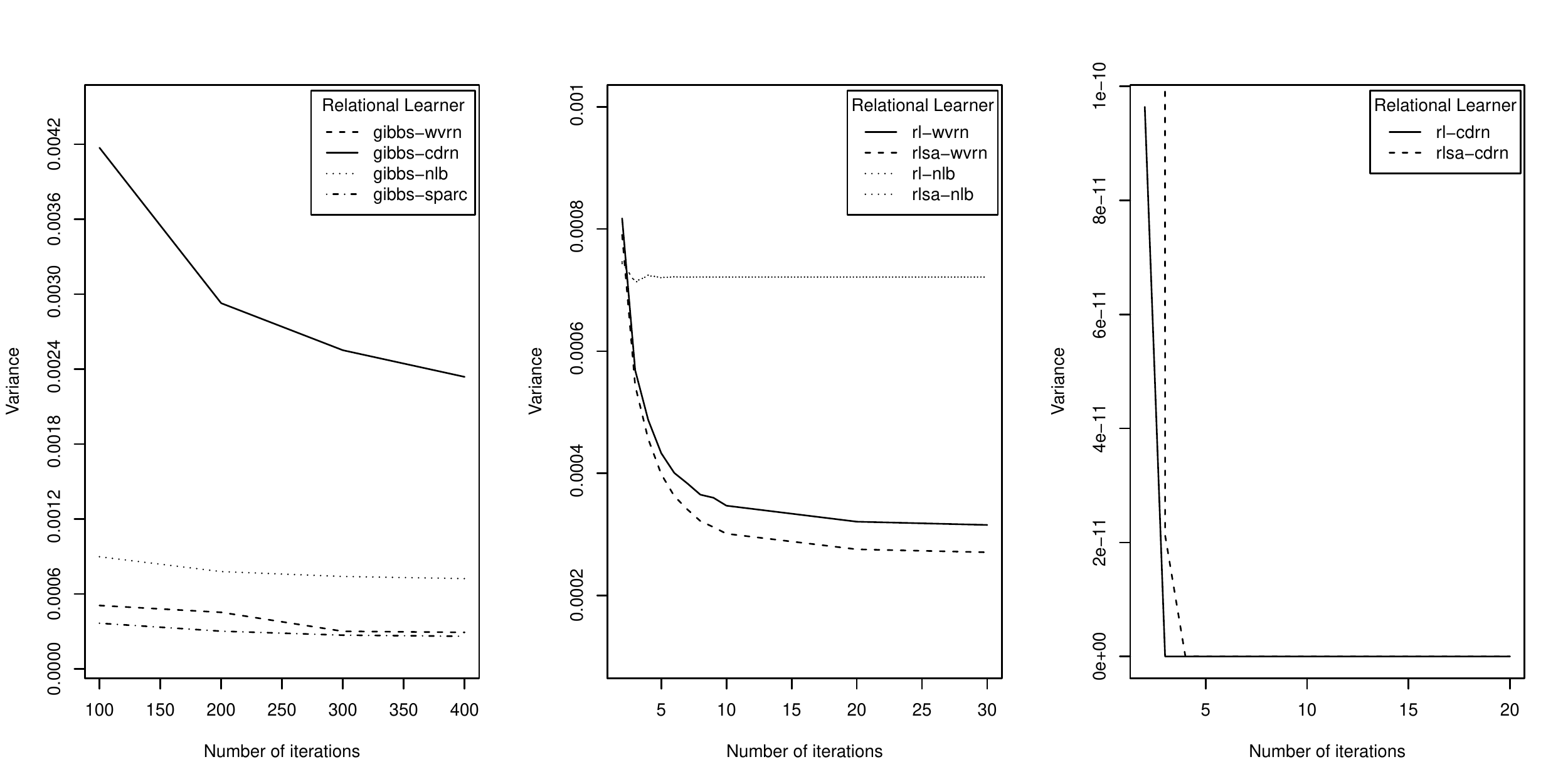}
\caption{\label{sensAn} The figures show the results of the sensitivity analysis of some of the relational learners. }
\end{figure}

When CI methods iteratively classify the nodes in a network, a smoothing effect of the churn influence may occur.
Before benchmarking the methods along all available datasets, we investigate this effect for each of the relational learners using one of the datasets.
Figure \ref{sensAn} shows the variance of the predicted churn scores as a function of the number of iterations in the collective inference method for various relational learners. 
This sensitivity analysis shows that the variation decreases very rapidly as the number of iterations increases.
This means that, with the default settings of each CI method, the variation of the final scores is very low and consequently there is very little distinction between churners and non-churners.
In particular, for the iterative classification CI, the scores stabilize after only a few iterations.
It can be seen from the figure, that for some relational classifiers, such as WVRN, this decline in variance happens faster than for the others.
We note that the CI spreading activation was not included, as it already contains an early stopping mechanism to compensate for this kind of smoothing effect.
Based on the sensitivity analysis, it was decided to implement the same kind of mechanism for early stopping in the other CIs.

Overall, we consider four relational classifiers and five collective inference methods, which, together with the option of not applying a CI, results in a total of two dozen unique combinations of methods, or 24 relational learners.

\subsection{Performance Measures}\label{perfM}

We next discuss two commonly used measures -- lift and AUC -- in addition to two more recent ones -- the Maximum Profit and the Expected Maximum Profit measures.
The first two measures are applicable in classification problems, whereas the last two are specifically developed for churn prediction models.

Lift is a commonly used performance measure for customer churn \citep{hung2006applying, tuffery2011data,verbeke2014social}.  
It compares the ratio of churners in a fraction of customers with the highest predicted probabilities to the ratio of churners in the actual customer base.
Thereby, it represents how much better a prediction model is at identifying churners, compared to a random sample of customers.
The top decile lift, considering the top $10\%$ of the customer base, is typically used.
However, this choice of the top fraction is arbitrary and does not necessarily reflect the needs of the company, especially if the customer base is large, as is the case for some of the datasets in this study, see table \ref{datasets}.  
Including too many customers in a retention campaign does not only increase the cost of the campaign but also increases the risk of offering promotions to customers who have no intention of churning.
Therefore, it is more meaningful to consider smaller fractions to identify only those who are most likely to churn.
To compensate for the variation in number of observations in our datasets we choose to use lift at $0.5\%$ and $5\%$.
As a result, the lift measure proves meaningful for both large and small organizations.

Another widely used and well established performance measure is the receiver operating characteristic  curve (ROC) and its corresponding AUC value \citep{fawcett2006introduction}.
AUC is typically a number between 0.5 and 1 and encapsulates the trade-off between the true and false positive rates. 
It can be interpreted as the probability that a randomly chosen churner is ranked higher than a randomly chosen non-churner. 
Recent studies have suggested Hand's H-measure as a coherent alternative to the AUC measure \citep{hand2009measuring} but because of evidence of correlation between these two  \citep{lessmann2015benchmarking}, we will only include the widely used AUC.

In a business setting, CCP modeling is usually conducted to decide which customers to target in a retention campaign.
As such, the specific requirements of the campaign should be taken into account when evaluating the models, to accurately measure the anticipated profit.
The maximum profit measure \citep[MP;][]{verbeke2012new} was developed with this objective. 
Based on the total profit of a retention campaign, as proposed by \citet{neslin2006defection}, the average classification profit for customer churn is defined as
\[
P^{ccp}(t;\gamma,CLV,\delta,\phi)=CLV(\gamma(1-\delta)-\phi)\cdot \pi_0 F_0(t)-CLV(\delta+\phi)\pi_1F_1(t)
\]
with $\eta$ the fraction of customers that are targeted, $\gamma$ the probability that a customer accepts the incentive, $CLV$ the average customer lifetime value, $\delta$ the cost of the incentive, $\pi_0$ the base churn rate and $\pi_1$ the base non-churn rate, $F_0(t)$ and $F_1(t)$ the cumulative density functions for churn and non-churn, respectively, given the cut-off $t$, $\lambda$ the percentage of churners within the targeted fraction, and $\phi$ the cost of contacting the customer.
The maximum profit ($MP$) measure is then defined as 
\[
MP^{ccp}=\max_{\forall t} P^{ccp}(t;\eta,\gamma,CLV,\delta,\phi)
\]
and represents the fraction of customers that should be targeted for a campaign to achieve maximum profit.
As many of the parameters are not always known, a corresponding expected maximum profit ($EMP$) measure \citep{verbraken2013novel} is proposed. 
It is given by
\[
EMP^{ccp}=\int_{\gamma}P^{ccp}T(\gamma);\gamma, CLV,\delta,\phi)\cdot h(\gamma)d\gamma
\]
where $T$ is the optimal cut-off value for a given $\gamma$ and $h(\gamma)$ is the probability density function for $\gamma$, chosen as the beta distribution as in the $H$-measure.

The maximum profit measure has advantages over both the lift measure and AUC.
Firstly, the fractions chosen for the lift measure are arbitrary and take neither costs nor benefits into account when measuring the performance. 
While both lift and AUC are universal and applicable to any classification problem, MP and EMP are the only measures that are specifically designed to take into account the requirements of a CCP model, which is the campaign itself, its costs, and its expected return.
Here, we choose to use the EMP measure because it is more robust in terms of the inherent stochastic nature of costs and benefits in churn management.

\section{Experimental Design}
\subsection{Data Description and Preparation}\label{subsec:datadescr}
\begin{table}[t]\footnotesize
\caption{\label{datasets} Dataset Description}
\begin{center}
\scalebox{0.9}{
\begin{tabular}{lp{2cm}p{0.8cm}p{1.4cm}p{1.9cm}p{1.5cm}l}
\hline
ID&Origin&Year&Customers& Churn Rate&Sparsity\\ \hline
BC1&Belgium&2010 &$1.41 \cdot 10^6$&$4.4\%$&$7.93\cdot 10^{-7}$\\
BC2&Belgium&2010&$1.21 \cdot 10^6$&$0.84\%$&$2.20\cdot 10^{-6}$\\
GD1&North America&2015&$1.57 \cdot 10^6$&$0.71\%$&$3.14\cdot 10^{-6}$\\
GD2&North America&2015&$1.32 \cdot 10^6$&$2.5\%$&$1.69\cdot 10^{-6}$\\
BP1&Europe&2009 &$4.33 \cdot 10^6$&$8.5\%$&$9.42\cdot 10^{-7}$\\
BP2&Europe&2008&$4.52 \cdot 10^6$&$3.5\%$&$9.44\cdot 10^{-7}$\\
MV&Belgium&2012&$1.70 \cdot 10^5$&$8.3\%$&$1.86\cdot 10^{-5}$\\
IS&Iceland&2015&$9.36 \cdot 10^4$&$2.3\%$&$1.04\cdot 10^{-4}$\\ \hline
\end{tabular}}
\end{center}
\end{table}

Following collection, eight distinct CDR datasets were analyzed in order to provide answers to the research questions in Table \ref{T:RQ}.
Table \ref{datasets} summarizes the main features of the datasets.
Most of the data come from telcos in Europe, while some have their origins in North America.

All datasets contain six consecutive months of cell phone usage data for the customers of the respective company, with information about the time, date and length of phone calls, and in some cases text messages and data roaming.
The oldest records are from 2008 and the most recent ones are from 2016.
Before building networks and churn labels, the datasets are all pre-processed in the same way. 
Only in-network phone calls are considered, and -- based on data exploration -- those lasting less than four seconds are disregarded.
Filtering the CDR data this way is a common approach in this type of research \citep{dasgupta2008social}, firstly because call records from competing telcos are usually not available and, as a result, neither is information about customers outside the network and secondly, shorter phone calls are regarded as unintentional.

As Table \ref{datasets} shows, the datasets vary greatly in number of customers, which ranges from less than a hundred thousand to over four million.
Actual churn dates are not available for all the datasets so, for the sake of consistency, we use the following definitions for churners and churn dates, based on literature  \citep{backiel2014mining, verbeke2014social}:
\begin{itemize}
\item[]\textit{Churner}: A customer with no perceivable activity for 30 consecutive days.
\item[]\textit{Churndate}: The first day of the 30 consecutive inactive days.
\end{itemize}
These definitions are appropriate for our purposes, because reduced activity is not only undesirable, but tracing and detecting churn rapidly, in order to offer incentives to remain loyal before it is too late, is highly important.
Because of these definitions the last month is needed to build the churn labels and therefore excluded from the analyses.
To ensure comparability of the different types of models, the fifth month is always used as the test and prediction month.
As a result, the first four months of each dataset are used to build the models.
The fraction of churners in the customer base, or the churn rates, are shown in Table \ref{datasets} for the prediction month.

A final observation about the datasets concerns the sparsity of the networks.
Sparsity is a typical characteristic of social networks, since the number of people is often very high but each person is only connected to a limited fraction of the network.
We can define sparsity as the fraction of non-zero elements in a matrix, which means that the lower it is, the fewer edges there are in the network.
As a result, lower sparsity means that the customers are less connected within the network.
Table \ref{datasets} displays the sparsity of each dataset for the month used as train data.
\subsection{Definition of Timeline and Networks}
For each of the eight datasets, the call-detail records span six months which we refer to chronologically as $M1$, $M2$, $M3$, $M4$, $M5$, and $M6$.
This partition of the data into consecutive months is the foundation for the timeline of our experimental setup.
As mentioned above, the last month $M6$ is needed to create the churn labels and is therefore left out of the analyses, and $M5$ is used for prediction and validation, which leaves $M1$, $M2$, $M3$ and $M4$ for model building.

The CDR data is aggregated to construct the networks for the analyses, where several decisions have to be made about the definitions of edges and weights.
Some are based on literature and others on specific factors that we intend to empirically estimate.
To measure effects both within the network when the call was made, and in the network over a longer period,
we look at two time periods, short-term and long-term, defined as one and three months, respectively. 
Thus, we build separate networks by aggregating information from CDRs over either one or three months.
The weights of the edges are computed in two different ways: using the total number of phone calls between two customers, or the aggregated length --measured in seconds-- of all phone calls between them, in a given time period. Both of these definitions are common in the literature \citep{baras2014effect,kim2014improved}.

Due to insights about the recency of edges in \citet{raeder2011predictors} and \citet{kusuma2013combining}, greater importance is ascribed to more recent connections.  
We use Equation \ref{formula} to apply decay to the weights, with time measured in weeks and the decay constant set to give links three months in the past some effect, but links that were made over a year before churning no effect. 
This selection is made based on exploration of the data and expert knowledge.
Because the factors `time span' and `weights' have two levels each, there are four different combinations of the two factors. 

The relational learners and the non-relational classifiers require different setups for conducting the experiments, as we describe below.
\begin{figure}
\centering
\subfloat[Relational Learners]{\includegraphics[width=11cm]{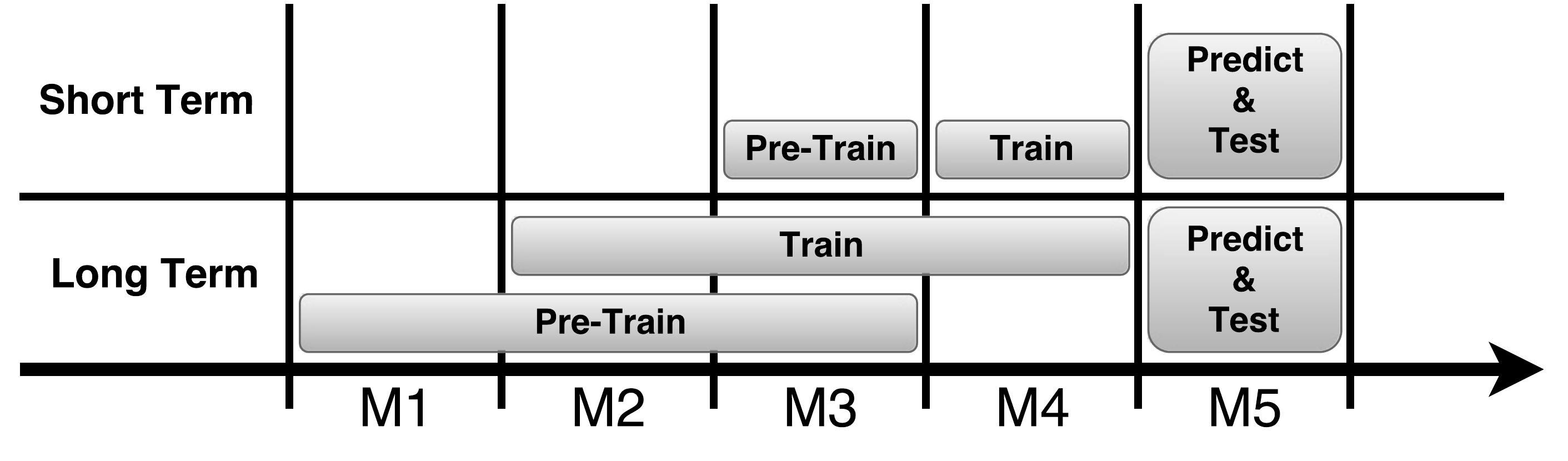}\label{F:timelineRL}}
\hfill
\subfloat[Non-Relational Classifiers]{\includegraphics[width=11cm]{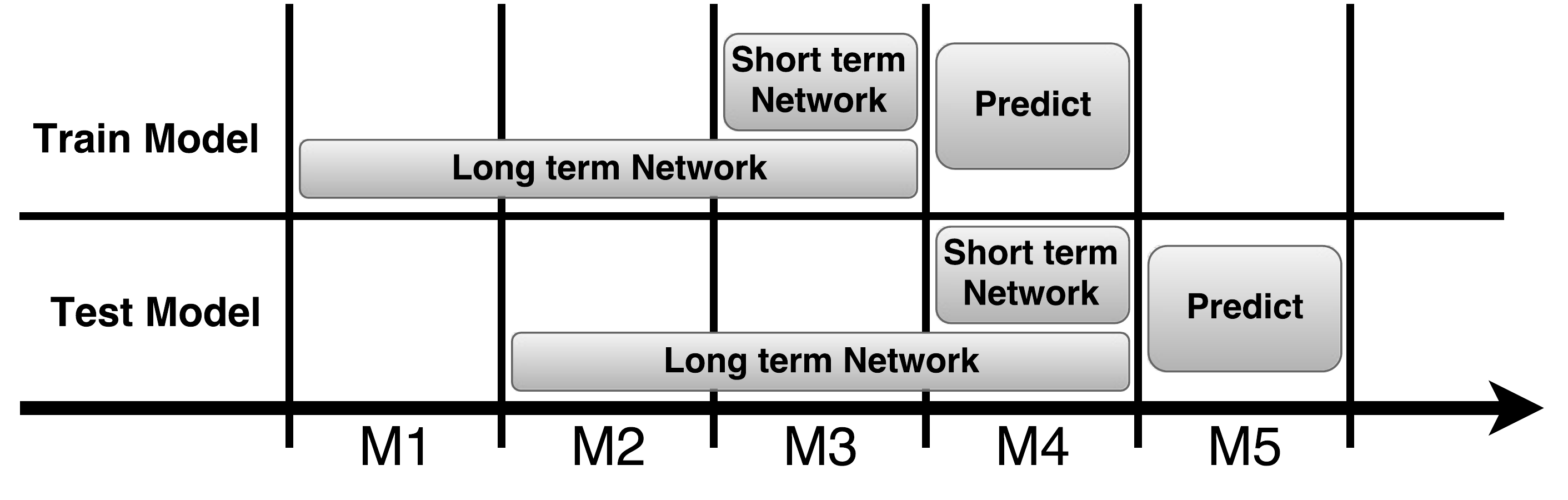}\label{F:timelineNRC}}
\caption{\label{F:timeline}The figures show how the datasets are split up by month to build the networks for pre-training, training and predicting in the short- and long-term settings for the relational learners and for out-of-time testing for the non-relational classifiers.}
\end{figure}
\subsubsection{Setup for Relational Learners}
In case of relational learners, our objective is to rank all two dozen of them, as well as evaluate the RC and CI separately.
We implement the time aspect of CCP as in \citet{verbeke2014social}, that is, we train the methods at a specific time, $t$, where all labels are known, and use the resulting scores as the estimated labels at time $t+1$, which in this case equals $M5$.
This means that we assume that $\mathcal{V}^K=\emptyset$ at time $t+1$ before the analyses are undertaken.
 
As described in Appendix \ref{App:RL}, two of the relational learners, CDRN and NLB, need a pre-training period, defined as having the same length as the corresponding training period, but starting one month prior as in \citet{verbeke2014social}.   
In summary, eight networks are built for the analysis of each dataset: 
for each weight representation, there are two short-term networks, $M3$ and $M4$, and two long-term networks, $M1$ to $M3$ and $M2$ to $M4$.
This is depicted in Figure \ref{F:timelineRL}.

Overall, each RL will produce four sets of scores for each of the eight datasets, or a total of 32 scores.

As is evident from Table \ref{datasets} there is severe imbalance in the distribution of the two classes, churners and non-churners.
Because of the networked nature of the data, commonly used sampling techniques, such as over- and under-sampling, are not applicable when using relational learners.
Since the goal of the learning is to rank the customers and not assign churn probabilities to them, 
we do not address this problem further.

\subsubsection{Setup for Non-Relational Classifiers}
The second part of the experimental setup is to build churn prediction models using non-relational classifiers with network features and scores from relational learners.
Thus, we try three combinations of features: network features only, RL scores only, and network features together with RL scores.
Common network features\footnote{Network features include degree, full, churn and non-churn; triangles, full, churn and non-churn; transitivity and link-based measures, described in subsection \ref{nrc}.} and various RFM features, such as number of days since last phone call, number of phone calls and total length of phone calls in the last 30, 60 and 90 days, are extracted from the four network structures described above, in addition to RL scores.
For robustness, an Out-of-Time (OoT) experimental setup is used for this part, as is depicted in Figure \ref{F:timelineNRC}.
As before, churn labels of month $M5$ are used for validation, but when training the models, churn labels of month $M4$ are used.
Variables for the training dataset originate from months $M1$, $M2$, and $M3$.
Thus, models to predict churn in $M4$ are trained on data from months $M1$, $M2$, and $M3$ and subsequently the models are applied to the same variables from months $M2$, $M3$, and $M4$, to predict churn in month $M5$.
As mentioned above, the classifiers we use are logistic regression, artificial neural networks and random forests.
The random forest models are trained using 500 trees and one hidden layer is used in the neural networks since this is sufficient for the neural network to be a universal approximator. 
Other parameters such as the number of hidden units were tuned experimentally on a separate validation dataset. 
Because of the class imbalance, we applied oversampling to the training datasets before building classification models.

\section{Relational and Non-Relational Learner Effects}
In this section, we present and discuss the results of the experiments.
We follow the guidelines of \citep{demvsar2006statistical} to statistically compare classifiers and methods.
More precisely, to measure a difference in performance of various methods, a Friedman test is applied to the average rank in performance of the methods. 
A low p-value means that the null hypothesis of no significant difference between the methods can be rejected.
If the null hypothesis is rejected, a post-hoc Nemenyi test is applied to explore which methods diverge from the rest \citep{verbeke2012new}.
In other cases, we turn to the non-parametric Kruskal-Wallis test, to investigate differences in performance, where a low p-value means that the null hypothesis of the samples originating from the same distribution can be rejected.
We use a $95\%$ confidence level when reporting significance of results.

Appendix \ref{app:abb} has a table where all abbreviations are explained.

\subsection{Relational Learners}
The first four research questions in Table \ref{T:RQ} involve comparing and ranking all relational learners in addition to the five collective inference methods and four relational classifiers.

\begin{figure}
\centering
\includegraphics[scale=0.4]{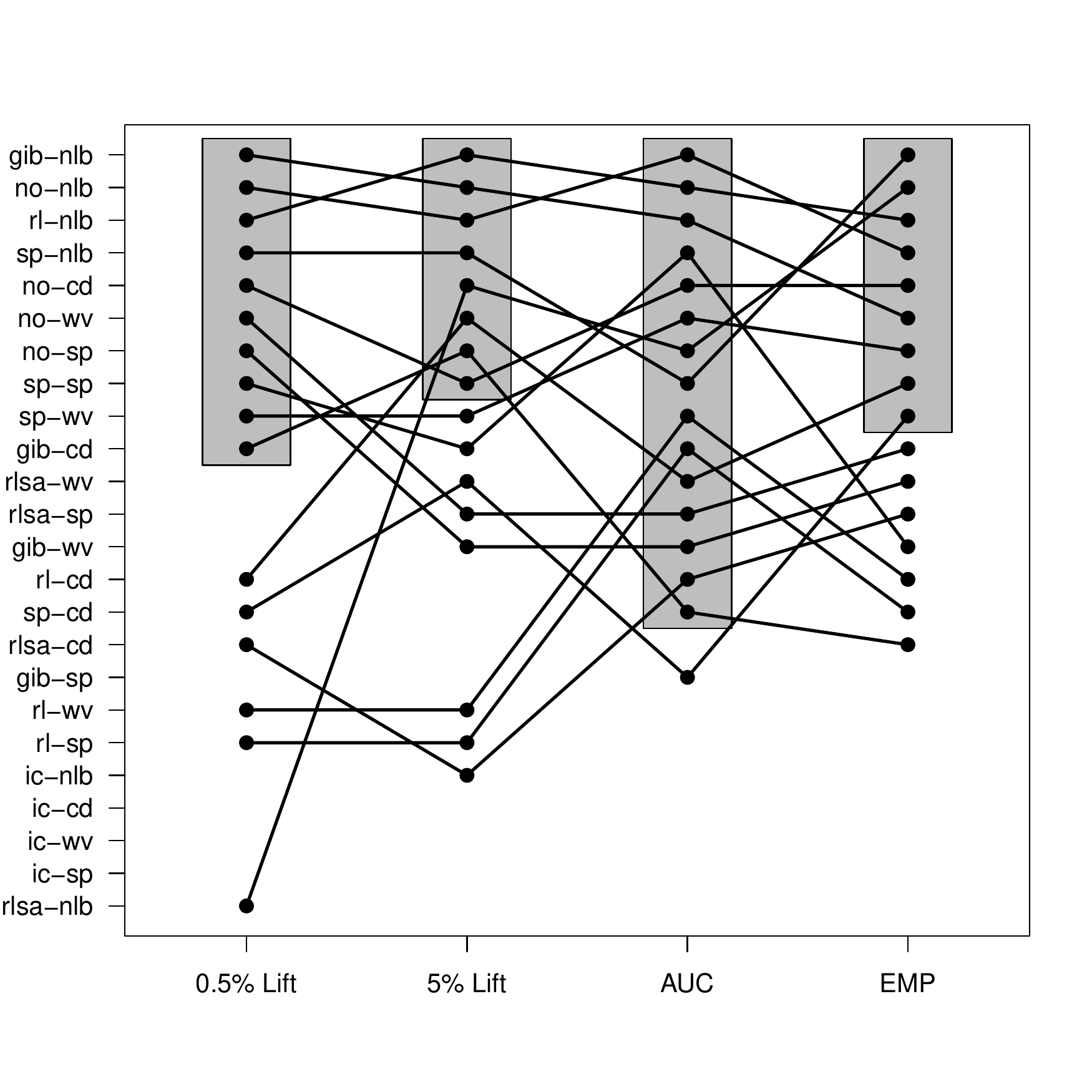}
\caption{\label{rank} The figure shows the relative ranks, from best, at the top, to worst, at the bottom, as measured by the four performance measures. The gray boxes represent the methods which are not significantly different at the $95\%$ confidence level.  Lines for the methods which performed significantly worse by all measures were not drawn. }
\end{figure}

Friedman tests applied to the performance of the relational learners results in p-values of less than 0.01 for all performance measures, which means there is a significant difference in performance between at least some of them.
A post-hoc Nemenyi test established the results in Figure \ref{rank}, where the methods in the gray boxes are those that do not differ significantly from the best performing one according to the four measures, $0.5\%$ lift, $5\%$ lift, AUC and EMP.  
When performance is measured by $0.5\%$ lift, 14 relational learners perform worse than the best performing one, 16 when measured by $5\%$ lift, nine when measured by AUC, and 15 according to EMP.
Moreover, four of the relational learners consistently perform the best according to all four performance measures, and nine relational learners perform worse than the best performing method for all measures. 
From the figure, it is evident that there is a significant difference between some of the methods; and it is also evident that the same methods tend to consistently outperform some of the others.
Noticeably, relational learners which have the relational classifier network-only link-based classifier (NLB) tend to perform better, except when combined with the collective inference method iterative classification (IC).
Clearly, there is a significant difference between good and bad methods.

\begin{figure}
\centering
 \hspace*{\fill}%
\subfloat[Differences of Relational Classifiers]{\includegraphics[scale=0.39]{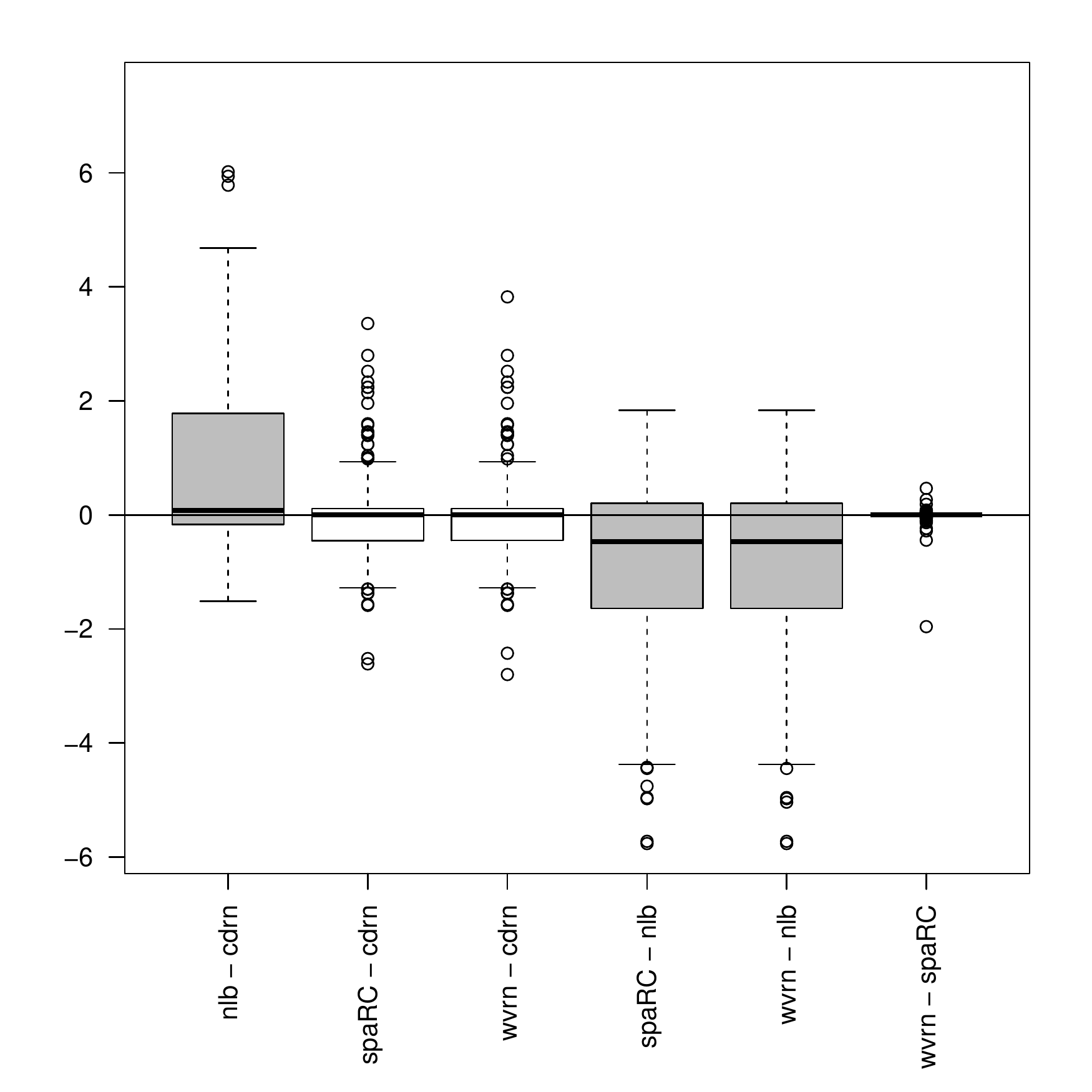} \label{F:diffRC}} \hfill
\subfloat[Differences of Collective Inference Methods]{\includegraphics[scale=0.39]{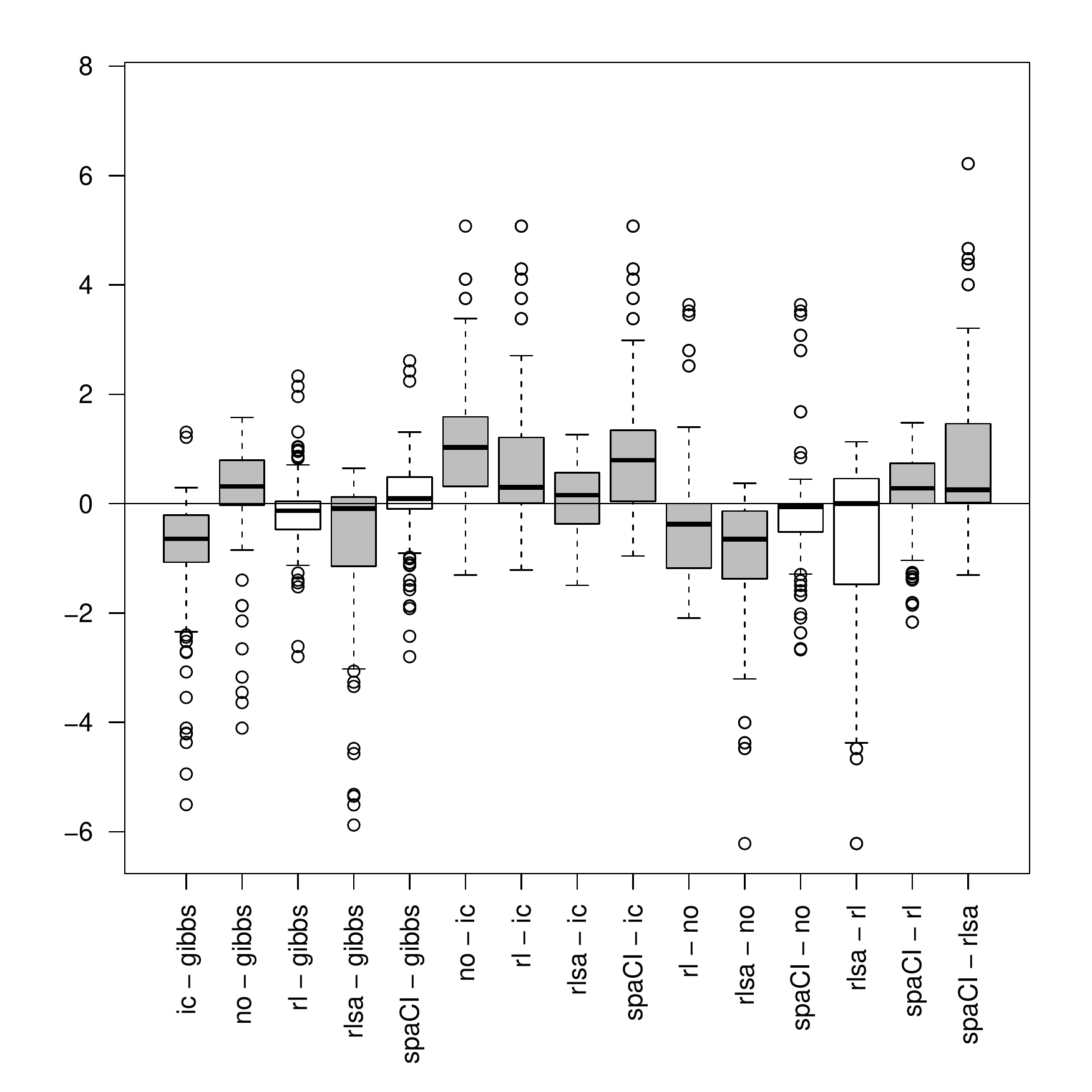}\label{F:diffCI}} 
 \hspace*{\fill}%
\caption{\label{F:diffHalf} The figures show pairwise differences in performance measured by $0.5\%$ lift of the four relational classifiers (left) and the six collective inference methods (right). Gray- and white- colored boxes represent differences that are and are not significant at the $95\%$ confidence level, respectively.}
\end{figure}

We further investigate this difference by applying a Friedman test to the four relational classifiers.
The p-values for all performance measures are less than 0.05 and therefore we continue to explore the pairwise differences between the classifiers.
The results can be seen in Figure \ref{F:diffRC} which shows the difference in performance between all RC, measured by $0.5\%$ lift.
The gray boxes, which represent significant differences, show that the differences NLB-CDRN, spaRC-NLB and WVRN-NLB are all significant. 
This implies that NLB performs significantly better than all the other relational classifiers, and it is the only one that differs from the rest.
Similar figures for the other performance measures show the same results.
The superiority of NLB can be explained by how comprehensive it is, as it considers the whole network before inferring labels. 
In a pre-training step, this classifier builds a logistic regression model for the nodes in the network, using their link-based measures.
This model is then used in the inferencing process, and because it already has internalized the information from the whole network it is able to make more accurate predictions.
In contrast, the other relational classifiers calculate a score based on each node's neighborhood, without any pre-training and without taking into account the rest of the network.

The collective inference methods are tested in the same way.
Again the Friedman test results in low enough p-values   and the results of a post-hoc analysis of the differences can be seen in Figure \ref{F:diffCI}. 
The results in the figure are based on the $0.5\%$ lift, but performance according to the other measures shows the same behavior.
As the figure shows, IC always performs significantly worse than the other methods, and not applying CI performs significantly better than applying any of the CIs.
A probable explanation for why IC performs worse than the rest is that, in each step of the iteration, it assigns churn labels and not probabilities or scores like the other methods.
By making such absolute decisions, a significant amount of information is lost, and there is less flexibility when inferencing in subsequent steps; therefore the prediction-making process becomes less accurate on the whole.
The sensitivity analysis in subsection \ref{secRL} even reveals that the assigned labels do not change after the first few iterations.

Finally, we test whether there is a difference in performance between methods with CI and without CI using the parametric Kruskal-Wallis test.
The p-value for each performance measure was less than 0.001, meaning that there is strong statistical evidence of the difference between methods that apply CI and methods that do not.
Further investigation shows that methods without CI performed better.
Although this may be a contradictory result to previous findings on the performance of collective inference methods \citep{jensen2004collective,sen2008collective}, there are various possible explanations for this behavior.
In fact, as we have already shown, the variation in churn scores decreases very rapidly when CIs are applied, thus decreasing the predictive performance.
The call networks in this specific problem setting are very large and sparse, and with few churn signals.  
As a result, when the CIs start spreading the `churn influence' it gets too diluted and in the end the signals are not strong enough or too similar to the non-signals to be meaningful.
This also seems to suggest that `churn influence' does not spread much beyond a person's first neighborhood, meaning that only the friends of a churner are affected by it and the friends of friends much less so.

To conclude, we have successfully answered the first four research questions.  
Figure \ref{rank} provides a ranking of the performance of the relational learners (RQ1).
Our results show that NLB is the best performing relational classifier (RQ2) and that there is no best performing collective inference method although IC performs the worst (RQ3).
Finally, we can conclude that collective inference methods do not improve the performance of relational classifiers (RQ4). 

\subsection{Non-Relational Classifiers}
\begin{figure}
\centering
\includegraphics[scale=0.70]{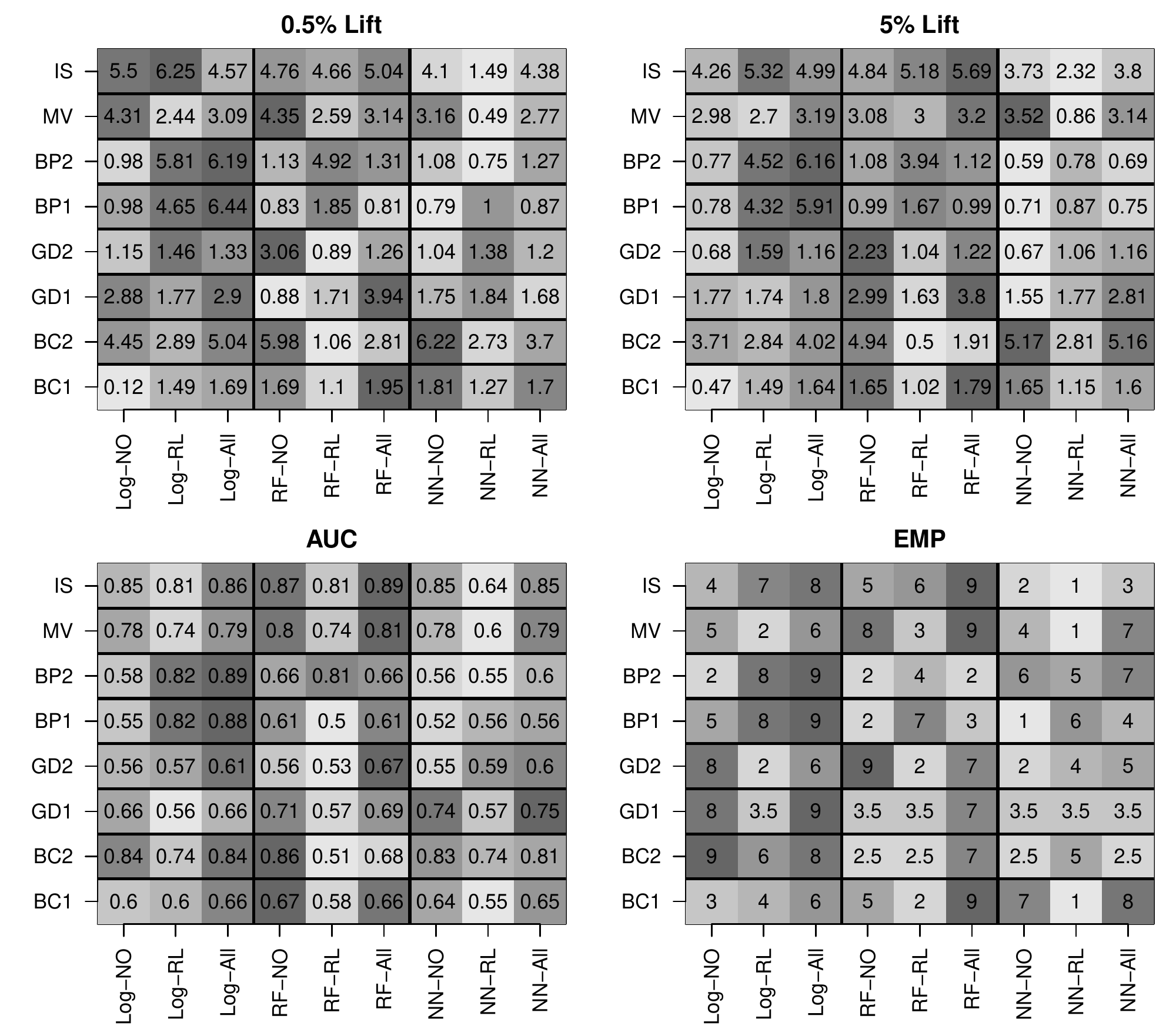}
\caption{\label{NRC_comparison} The figures compare the performance of each of the three non-relational classifiers with the three subsets of variables for the eight datasets and the four performance measures.  Each row represents one dataset. In the first three columns on the left are the logistic regression models (Log), in the three columns in the middle are the random forests models (RF) and in the last three columns on the right are the neural networks models (NN).  Each classifier is used to build a model using network variables only (NO), RL scores only (RL), and network variables together with RL scores (All). The figures show the the measurement values --or the actual rank in the case of EMP-- of each combination and a color coded ranking of the nine models for each dataset, with a darker color representing a better performance. }
\end{figure}
The second part of the analysis aims to answer research question RQ5 about non-relational classifiers.

A Friedman test is applied to the results of churn prediction models built with the binary classifiers logistic regression, random forests and neural networks; using as attributes, network features only, RL scores only, and both of these combined.
The p-value for all performance measures was less than 0.01 except for lift at $0.5\% $, where the p-value equalled $0.17$. 
A comparison of the performance of the models can be seen in Figure \ref{NRC_comparison} which shows the relative ranking of each combination of classifier and feature set for each of the eight datasets.
In the figure, a lighter gray color means a worse performance and a darker gray signifies a better performance. 
In addition, the measurement values are shown for the two lift measures and AUC but the actual ranking for EMP, since these values were too varied to fit adequately in the figure.
Subsequent analysis reveals that the models with both types of variables perform significantly better than models with only one of them.
This is evident by the fact that for each of the three triples in every line of the four figures, the column on the right, which represents models built using both sets of features, tends to be darker than the other two columns.
Although that is not the case for some of the datasets,
it is the overall effect.

Finally, no significant difference is found between the performance of models built with either RL scores or network variables; only models with a combination of the two features performed significantly better.
This illustrates that adding features increases the model performance and that relational learner scores do possess valuable knowledge that is not captured by network features alone.
This is further confirmed by the fact that the scores were almost never correlated with the network features, although there were often correlations within each set of features.
However, scores produced when applying the CI Gibbs sampling were never correlated with any other features and not even amongst themselves, which was a common behavior for the other RL scores.

We can conclude, based on our results, that models with a combination of network features and relational learner scores perform best (RQ5).

\subsection{Comparison of RL and NRC}
Regarding research question RQ6, when comparing the predictive performance of relational learners and non-relational classifiers, we apply the Kruskal-Wallis test which results in p-values of less than 0.01 for all performance measures.
Further exploration reveals that the non-relational classifiers, with any set of features, perform significantly better than the relational learners.
This result is understandable, because relational learners take into account fewer network properties than the non-relational classifiers do.
The former approach, which simulates the effect people have on others further away in the network and takes into account how churners in the network influence each other, is therefore a valuable addition to the latter approach.
It also shows that, for the purpose of accurate churn predictions, relational learners are better when used in combination with non-relational classifiers.

\begin{table}
\caption{\label{T:pValues}P-values from the Kruskal-Wallis test when comparing the difference in performance between each relational learner listed in the table and all non-relational classifiers.}
\centering
\scalebox{0.78}{
\begin{tabular}{lrrrr}
\hline
&\multicolumn{4}{c}{Performance Measure}\\ \cline{2-5}
Relational Learner&$0.5\%$ lift& $5\%$ lift& AUC& EMP\\ \hline
gibbs-nlb&0.47&0.65&$<0.001$&0.70\\
no-nlb&0.98&0.50&$<0.001$&0.64\\
rl-nlb&0.60&0.34&$<0.001$&0.81\\
spa-nlb&0.60&0.70&$<0.001$&0.81\\
no-cdrn&0.004&$<0.001$&$<0.001$&0.22\\
no-wvrn&0.003&$<0.001$&$<0.001$&0.16\\
no-spaRC&0.003&$<0.001$&$<0.001$&0.16\\
spa-sparc&0.002&$<0.001$&$<0.001$&0.071\\
\hline
\end{tabular}} 
\end{table} 

We note that these results hold for the cumulative performance of the relational learners.
When considered individually, it is evident that not all of the best performing relational learners are outperformed by the non-relational classifiers. 
This is verified using the Kruskal-Wallis test to compare the performance of each relational learner, that appears in the gray box for lift at $5\%$ in Figure \ref{rank}, to the overall performance of the three non-relational classifiers.
The resulting p-values for the four performance measures can be seen in Table \ref{T:pValues}.
The table shows that the first four relational learners are only outperformed by the non-relational classifiers when performance is measured using AUC, but not using any of the other three performance measures.
The last four relational learners are all outperformed by the non-relational classifiers, except when measured using EMP.
 
\begin{figure}
\centering
 \hspace*{\fill}%
\subfloat[Relational Learners]{\includegraphics[scale=0.19]{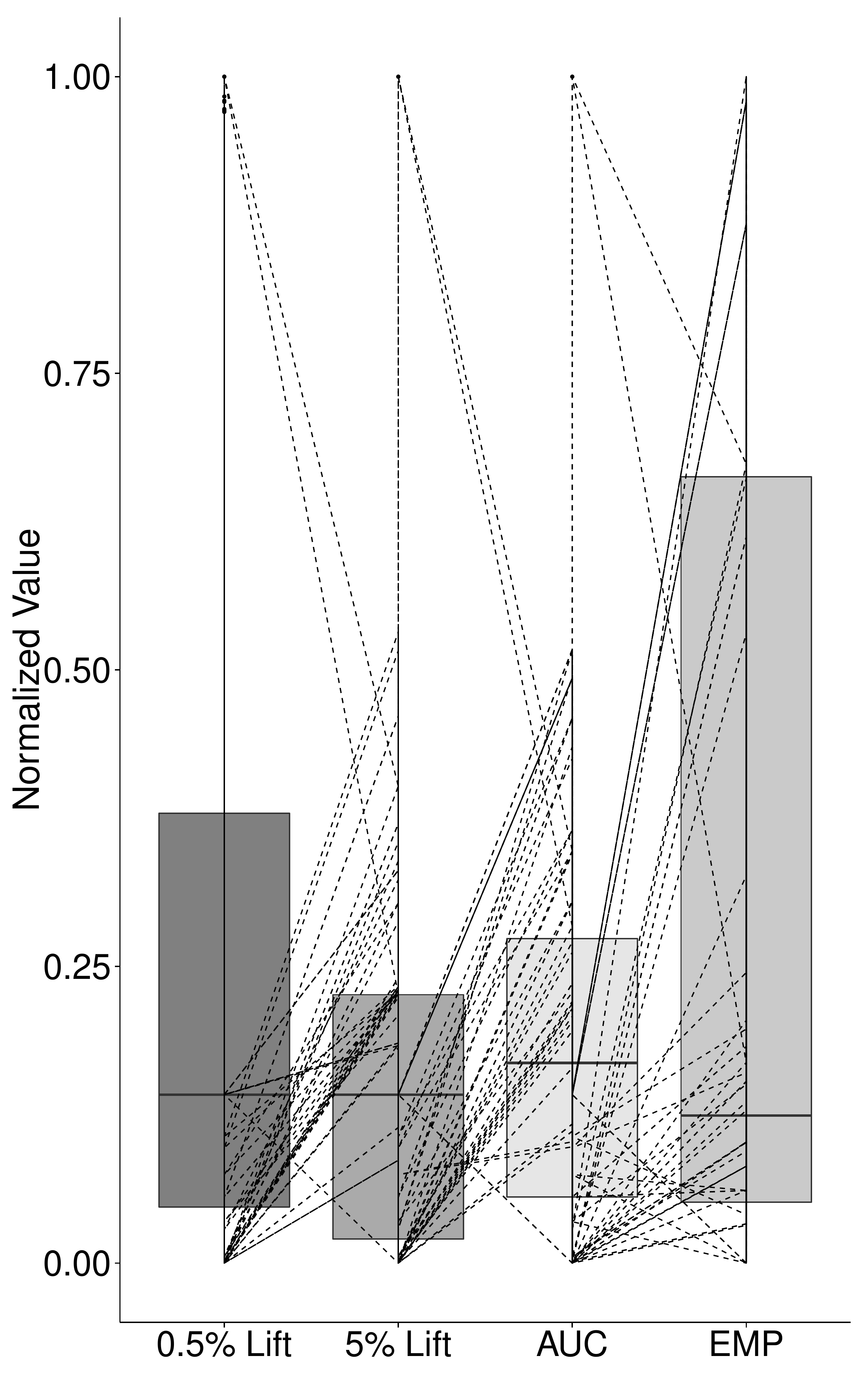}\label{BoxScatterRL}} \hfill
\subfloat[Non-Relational Classifiers]{\includegraphics[scale=0.19]{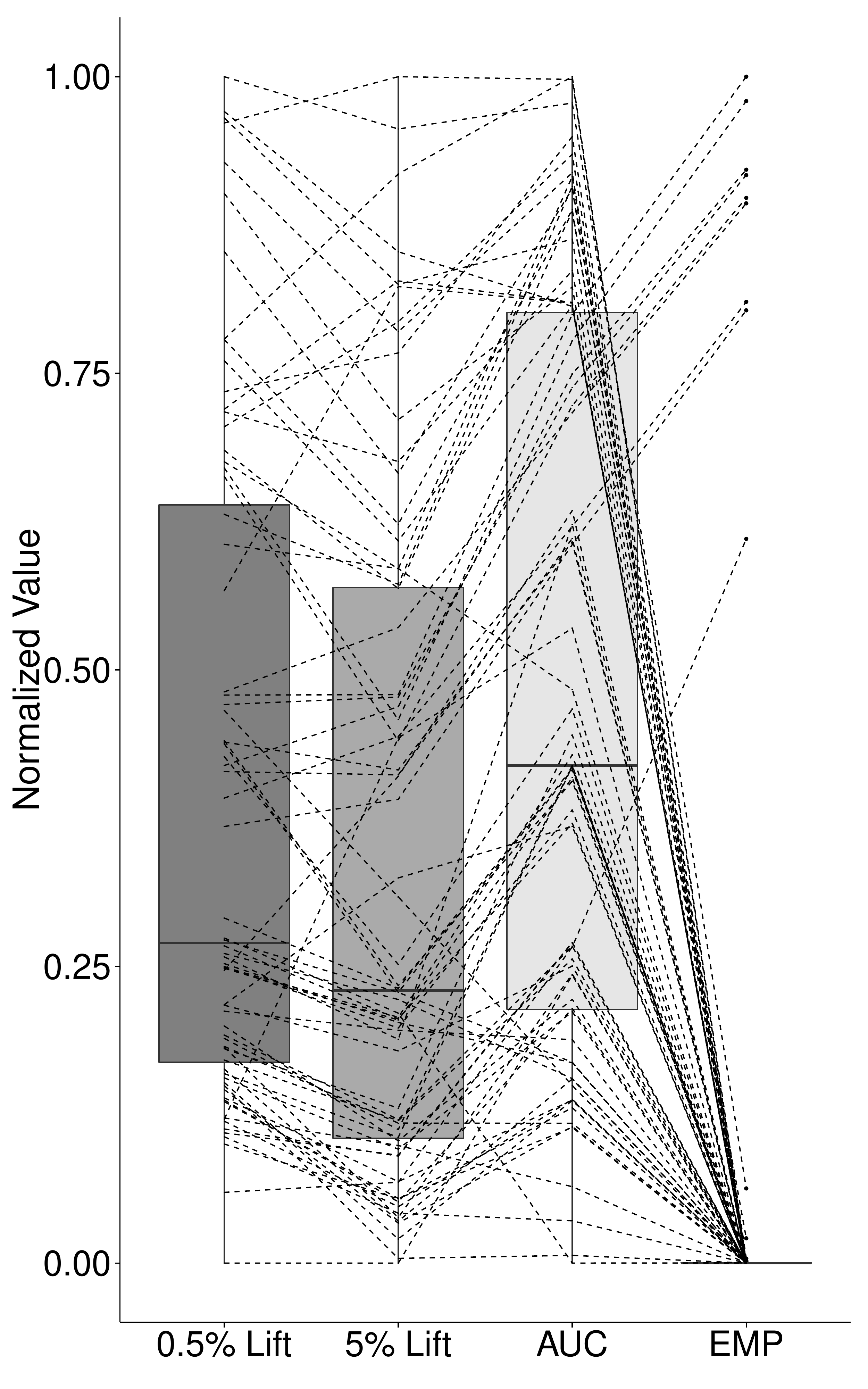}\label{BoxScatterNRC}} 
 \hspace*{\fill}%
\caption{\label{BoxScatter}The figures show, for the relational learners (left) and for the non-relational classifiers (right), the distribution of the four performance measures (boxplots) and the correlations between them (dotted lines).}
\end{figure}

We conclude our discussion of relational learners and non-relational classifiers by looking at their distribution and correlations, see Figure \ref{BoxScatter}.  
For the two approaches, we normalize the values of each performance measure in order to make them comparable and present their distributions with box-plots.
Subsequently, measurements of the same model are connected, as shown by the dotted lines.
The figure thus illustrates a combination of box- and scatter- plots with correlations.
The distributions of the measures vary, as the box plots show, and the values are also not very correlated, as illustrated by how frequently the dotted lines connecting the measurements intersect.
Since all the measures give consistent results, this is a good indication of the robustness of our findings.
Note that in Figure \ref{BoxScatterNRC}, most of the values of the EMP measure are very small, whereas one of the datasets resulted in values that were much higher (by a factor of $10^6$).
This explains why there is a cluster around 0 in the figure.

To answer the final research question, we assert that overall, non-relational classifiers perform better than relational learners, although a few of them are just as good at predicting churn as the non-relational methods. 

\section{Effects of Network Construction}\label{sec:expost}
As mentioned, research on the exact impact of network construction on models and findings in the telecommunication industry is scarce. 
In most studies, a network is built in a specific way, often without detailed supporting arguments or explanations \citep{dasgupta2008social}.
A construction issue which is overlooked in such studies is that the findings could depend on how the network is defined, and might therefore change if the network is constructed differently by whoever is conducting the experiment. 
Clearly, such a situation would result in potentially incorrect conclusions, as the predictive capability of the model would not be affected by the sheer explanatory power of the data but by the decisions of the modeler.

Here, in a follow-up study, we explore this potential effect,
by building more than 500 networks with varying definitions of edges and weights and aggregating them in numerous ways. 
The detailed setup to arrive at the 500 networks will be explained in subsection \ref{subsec:defining}.
The number of networks can increase very rapidly when taking all possibilities into account, and as a result, it becomes computationally infeasible to compare all model combinations. 
Instead, we exploit the result of the benchmarking study in the previous section for `algorithm selection'.
Since we have shown that CIs do not improve the performance of RC, those RLs can be left out, and based on Table \ref{T:pValues} some RLs are just as good at predicting churn as NRC is.
Therefore, we have chosen one relational learners (NLB) as a proxy method to predict churn in the different types of networks to compare the performance of the resulting models.
As a result, we hope to guide further studies on how to optimize network architecture when predicting churn in telco.

\subsection{Defining the Network Architecture}\label{subsec:defining}
\begin{table}
\caption{\label{T:networkDef} Segmentation of Networks}
\centering
\scalebox{0.7}{
\begin{tabular}{cm{3.3cm}m{2.5cm}m{3.5cm}m{3.5cm}}
\hline
Day&Part-of-Week&Time-of-Day&Comb. Part-of-Week&Comb. Time-of-Day\\ \hline
Monday&Working days(WD)&Day(8-16)&$\frac{1}{2}$WD$+$WE&$\frac{1}{2}$Day$+$Evening\\
Tuesday&Weekend(WE)&Evening(16-24)&WD$+\frac{1}{2}$WE&Day$+\frac{1}{2}$Evening\\
Wednesday&&Night(0-8)&$\frac{1}{3}$WD$+$WE&$\frac{1}{3}$Day$+$Evening\\
Thursday&&&WD$+\frac{1}{3}$WE&Day$+\frac{1}{3}$Evening\\
Friday&&&&\\
Saturday&&&&\\
Sunday&&&&\\ \hline
\end{tabular}}
\end{table}
For this follow-up study, we select the BC2 dataset as it represents an `average' among all the  datasets, both in size and sparsity.

As before, predictions are made for month $M5$ and, to reduce complexity, the relational learner was applied on one time frame -- the short-term time frame which spans one month.
Since NLB needs a pre-training period as well, months $M3$ and $M4$ are used for the pre-training and training, respectively.
As in section \ref{subsec:datadescr}, phone calls lasting less than four seconds are removed.
The edges of the network are defined in three ways: incoming, representing all phone calls made to a customer; outgoing, the phone calls made by a customer; and undirectional, when the previous distinction is not made.
The weights of the edges are defined in four ways.  
These are the aggregated length of all phone calls, the total number of all phone calls, the normalized average of these two, and finally an indicator of whether a phone call was made or not.
We refer to these as length, count, average and binary.
Section \ref{subsec:networks} explicates how weights can be weighted in time with decay using equation \ref{formula}
to confer more importance to more recent phone calls.
This technique is also included in our network construction together with aggregating the weights normally.  We call these two levels `with decay' and `simple weights'.
For each network the combination of weights, directions and decay leads to 24 different combinations.
The CDR data hold information about the date and time of each phone call.  
We use this information to segment the networks, thus building separate networks for phone calls made on each day of the week, during working days (WD) and weekends (WE) and during the day, evening and night. 
Finally, we combine times of the day and parts of the week in various ways to build more networks.
Table \ref{T:networkDef} shows a summary of all the networks that were built in this way.

On a social level, a relationship between two people only exists when both call the other.
As a result, some studies \citep{dasgupta2008social,haenlein2013social} remove connections that were not reciprocal before applying SNA.
To include this possibility in our exploration, we repeat the whole setup with non-reciprocal calls removed.

In total over 500 networks were built for the one dataset, which gives an indication of the scalability problem associated with these experiments and why more datasets and more relational learners are not included.
The proxy classifier is subsequently applied to all the networks, and the performance of the resulting models compared.

\subsection{Results of Network Construction}
\begin{table}
\caption{\label{T:WholeAUCLift} Results of whole network measured in AUC and (lift at $0.5\%$)}
\centering
\scalebox{0.78}{
\begin{tabular}{|l|cccc|cccc|c|}
\hline
&Length&Count&Average&Binary&Length&Count&Average&Binary&\\ \hline
Whole&0.748(3.21)& 0.748(3.21)& 0.504(1.16)& 0.748(3.21) &0.755(3.21)& 0.748(3.21)& 0.743(3.05)& 0.748(3.21)&\multirow{3}{*}{\rotatebox[origin=c]{90}{Undirected}} \\
Time-of-Day comb&0.748(3.19) &0.749(3.19)& 0.744(2.87)& 0.777 (3.19)&0.765(3.19)& 0.748(3.19)& 0.713 (3.20)&0.777(3.19)& \\
Part-of-Week comb&0.557(1.30)& 0.557(1.30)& 0.519(1.12)& 0.557(1.30) &0.748(3.21)& 0.748(3.21)& 0.698(2.05)& 0.748(3.21)& \\ \hline
Whole&0.741(2.54)& 0.741(2.54)& 0.502(1.14)& 0.741 (2.54)&0.741(2.54)& 0.741(2.54)& 0.721 (2.36)&0.741(2.54)&\multirow{3}{*}{\rotatebox[origin=c]{90}{Outgoing}}\\
Time-of-Day comb&0.740(2.56)& 0.740(2.56) &0.688 (1.93)&0.740(2.56) &0.743(2.56) &0.741(2.56)& 0.643(1.69) &0.740(2.56)& \\
Part-of-Week comb&0.535(1.26)& 0.535 (1.26)&0.510(1.12)& 0.535(1.26) &0.741(2.54)& 0.741(2.54)& 0.623(1.51)& 0.741(2.54)& \\ \hline
Whole&0.717(2.28) &0.717(2.26)& 0.703(2.22)& 0.717(2.28)&0.717(2.28)& 0.717(2.26)& 0.704(2.22) &0.717(2.28)& \multirow{3}{*}{\rotatebox[origin=c]{90}{Incoming}}\\
Time-of-Day comb&0.728(2.26)& 0.716(2.26)& 0.680(1.87)& 0.716(2.26) & 0.721(2.26)& 0.718(2.26)& 0.679(1.87)& 0.716(2.26)&\\
Part-of-Week comb&0.729(2.28)& 0.717(2.26)& 0.666(1.83)& 0.717(2.28) &0.723 (2.28)&0.717(2.26)& 0.666(1.83) &0.717(2.28) & \\ \hline
&\multicolumn{4}{c|}{Simple weights}&\multicolumn{4}{c|}{With decay}&\\ \hline
\end{tabular}}
\end{table}
We first discuss the results of the whole network where non-reciprocal edges have not been removed.

Table \ref{T:WholeAUCLift} shows the performance of the relational learner when predicting churn measured in AUC and lift at $0.5\%$.

For comprehensibility, we have chosen to show only three segmentations of the networks, namely the whole network, the best combination of part-of-week and the best combination of time-of-day.
The best combination of part-of-week is the fourth one in Table \ref{T:networkDef} or $
WD+\frac{1}{3}WE$ and the best combination of time-of-day is the first one in Table \ref{T:networkDef} which is $\frac{1}{2}\textrm{Day}+\textrm{Evening}$.
From Table \ref{T:WholeAUCLift} it is clear that the performance on the part-of-week combinations is lower than that on the other networks.
In addition, performance on the whole network tends to be slightly better than on the time-of-day combination network.
This result is the first interesting behavior that we can now extract. 
In some cases there is an increase in predictive capability by segmenting the network by part-of-week and also by time-of-day, with weekdays more important than weekends, and evenings more important than daytime. 
These results are consistent with the conclusion from our previous section: calling the close circle of the user (family and close friends) tends to be the most relevant factor when spreading churn influence. 
There is a greater chance that these calls occur in the evenings (after work) or during the weekdays, as, during the weekend, it is more likely that the users will be in the company of their close circle. 

Given that such a weighting scheme can be used to improve predictive capability, and that this is closely related to the nature of the problem, then a theoretical approach of first propagating the network influence over the whole network (the second best approach), later analyzing extracting on whom the influence is exerted, and finally weighting the networks to reflect this group might bring an improvement in predictive capabilities of a few percentage points.

When comparing the definition of the edges -- that is, the difference between performance on undirected, outgoing, and incoming networks -- it is clear the undirected outperforms the other two.
However, it is not as clear whether the outgoing or the incoming network is better.  
This result contradicts the findings of \cite{haenlein2013social}, where the outgoing edges are shown to have higher correlation with churn.
Given that our study uses a far larger number of mobile phone users (representing all calls in a given time frame) instead of a selected sample -- \citet{haenlein2013social} uses 3431 users --  we can venture that the directionality effect might occur only on some selected cliques, but in general it is only the presence of connectivity that is the main factor for allowing the spread of influence.

In addition, there seems to be a correlation between the edge definition and segmentation of the network, in that the performance of part-of-week network is higher for the incoming network and lower for the outgoing network, while the opposite holds for the time-of-day network.

Regarding the weights of the edges, the performance on the network with binary weights is almost consistently better or at least as good as on the networks with length and count, which have very similar performance.
This indicates that it does not matter whether length or counts are used for the weights and that the simplest variant, binary weights, will result in models that are just as efficient. 
On the average network, performance is always the worst. 
This behavior hints at the importance of the existence of connectivity rather than the intensity of the communication between two users. 
As mentioned several times, it is those in the close circle of the user who are more at risk of churn when the user churns, so the number, and particularly the length of the call would be a poor proxy of the intensity of the relationship (a couple that lives together would probably not make very long calls to each other consistently, for example) and,  as such, binary weights that reflect whether there is a connection or not would be sufficient to represent this relationship. 

Finally, we compare performance on networks with and without decay on the weights.
The networks with decay often slightly outperform the variant without decay, at least when performance is measured in AUC. 
This effect is not as clear in the table for lift at $0.5\%$.
This result, now consistent with \citet{haenlein2013social}, indicates that the influence of a churner is spread over the users that are in recent contact with the churner. 
This is quite obvious: the close circle is likely to be in permanent (so also recent) contact with the user, and users not in recent contact with the user would not have that effect.
Another interesting conclusion that can be extracted from this result is that the time frame for influencing is short, as our networks cover six months of calls at most.
\begin{table}
\caption{\label{T:recAUCLift}Results of reciprocal network measured in AUC and (lift at $0.5\%$).}
\centering
\scalebox{0.78}{
\begin{tabular}{|l|cccc|cccc|c|}
\hline
&Length&Count&Average&Binary&Length&Count&Average&Binary&\\ \hline
Whole&0.681(1.83) &0.681(1.83)& 0.676(1.81) &0.681(1.83) &0.681(1.83)& 0.681(1.83) &0.676(1.81)& 0.681(1.83)& \multirow{3}{*}{\rotatebox[origin=c]{90}{Undirected}} \\
Time-of-Day comb&0.681(1.85)&0.681(1.85)& 0.661 (1.71)&0.681(1.85) &0.681(1.85)& 0.681(1.85)& 0.661 (1.71)&0.681(1.85)& \\
Part-of-Week comb&0.681(1.83)&0.681(1.83) &0.651(1.67) &0.681(1.83) &0.682(1.83)& 0.681(1.83)& 0.651(1.67)& 0.681(1.83)& \\ \hline
Whole&0.681(1.83)& 0.681(1.83) &0.664(1.73)& 0.681(1.83) &0.683(1.83)& 0.681(1.83)& 0.664(1.73) &0.681 (1.83)&\multirow{3}{*}{\rotatebox[origin=c]{90}{Outgoing}}\\
Time-of-Day comb&0.681(1.85)& 0.680(1.85) &0.622(1.53)& 0.682(1.85) &0.680 (1.85)&0.680(1.85)& 0.622(1.53)& 0.682(1.85)& \\
Part-of-Week comb&0.681(1.83)& 0.681(1.83)& 0.606 (1.48)&0.681 (1.83)&0.682(1.83)& 0.681(1.83)& 0.606 (1.48)&0.681 (1.83)&\\ \hline
Whole&0.681(1.83)& 0.681(1.83)& 0.664(1.73)&0.681(1.87)&0.681(1.83)& 0.681(1.83) &0.664(1.73) &0.681(1.83)&\multirow{3}{*}{\rotatebox[origin=c]{90}{Incoming}}\\
Time-of-Day comb&0.681(1.87) &0.680(1.87)& 0.639 (1.61)&0.681(1.83)&0.680(1.87) &0.680(1.87) &0.639(1.61) &0.681(1.83)&\\
Part-of-Week comb&0.681(1.83) &0.681(1.83) &0.626(1.55) &0.681(1.83)&0.683(1.83)& 0.681(1.83)& 0.626(1.55) &0.681(1.83)&\\ \hline
&\multicolumn{4}{c|}{Simple weights}&\multicolumn{4}{c|}{With Decay}&\\
\hline
\end{tabular}}
\end{table}

We now turn to the results for networks with reciprocal edges only, which can be seen, measured in AUC and lift at $0.5\%$, in Table \ref{T:recAUCLift}.
Overall, the difference in performance between the types of edges, types of segmentation, types of weights and with and without decay is much less evident than above, and in most cases there is no difference.
It is possible to detect slightly better performance in the undirected network.
However, the clear difference between these results and the results in Table \ref{T:WholeAUCLift} is that, by removing the non-reciprocal edges, the performance has decreased significantly.

To conclude, the use of the best performing network -- that is, one which is constructed with binary weights, undirected and weighted in time -- brings about a significant increase in performance, so the modeler has a large responsibility to correctly define the best network for the problem that is being tackled, as failure to do so will result in less predictive capability and therefore less potential gains.

\section{Scientific and Managerial Insights}
Applying SNA for CCP in telco is a complex process where multiple factors need to be taken into consideration.
The results of the previous sections, however, can be used as guidelines on how to conduct this process in a practical way to achieve best performance.
In addition to being relevant for business users, the results also have important implications for academics because experiments as extensive as these ones, for SNA-based CCP in telco, do not have precedence in the literature.

Firstly, the follow-up study on the effect of network construction on model performance suggests that how the network is defined does indeed make a difference to the performance of the models that are applied to them.
Networks with undirected edges and binary weights show the highest performance.
In addition, more recent connections seem to matter more than older ones do.
This is a very meaningful result for the day-to-day business user since it implies that simpler networks are just as good as more complex ones.  
Therefore, less effort needs to be made when filtering data and building networks, thus simplifying the process. 
Furthermore, as more recent connections are more important, less data is needed.
Because of the lack of research on the effect of network construction on model performance, the consequences of our study are also important in an academic setting.  They could furthermore help in understanding human behavior when it comes to social ties and churn.
\begin{table}
\caption{\label{T:RQA} Research Questions with Answers. Performance is measured using four measures: lift at $0.5\%$ and $5\%$, AUC, and EMP as described in subsection \ref{perfM}.}
\centering
\scalebox{0.78}{
\begin{tabular}{|m{1cm}|m{0.5cm}m{17cm}|l|}
\hline
\multirow{4}{*}{\rotatebox[origin=c]{90}{\parbox[c]{3.5cm}{\centering Effect of  Relational Learners}}}&RQ1&Which relational learners perform statistically different from the rest when predicting customer churn?&See Figure \ref{rank}\\ \cline{2-4}
&RQ2&Do some relational classifiers perform statistically better than the others?&Yes\\\cline{2-4}
&RQ3&Is the performance of some collective inference methods statistically better than the others?&Yes\\\cline{2-4}
&RQ4&When predicting customer churn, do collective inference methods improve the predictive performance of relational classifiers?&No\\\hline \hline
\multirow{3}{*}{\rotatebox[origin=c]{90}{\parbox[r]{2.5cm}{\centering Combination of  RL and NRC}}}&RQ5&Which non-relational classifier model performs best when predicting churn?  A model built using network features only, a model with relational learners scores only, or a model which is built with a combination of both?&Both\\\cline{2-4}
&RQ6&Which model type performs better when predicting churn? Relational learners or non-relational classifiers?&NRC\\
&&&\\\hline
\end{tabular}}
\end{table}

Secondly, our extensive empirical comparison of the performance of relational learners and their combination with binary classifiers 
can be summarized with the six research questions and answers in
Table \ref{T:RQA}.
We compare all relational learners collectively, as well as their two components -- relational classifiers and collective inference methods.
Our tests reveal a small group of learners that perform better overall, and which are significantly better at predicting churn than the rest.
The classifier, network-only link-based classifier proposed by \citet{lu2003link} is the only one that significantly differs from the others, and it also proved to be the best performing one.
This classifier uses the whole network to build a logistic regression model with link-based features from the network.
Not only does it take into account information from a node's neighborhood when inferring a label, like the other classifiers, but also considers the joint information from the whole network.
In fact, it is very similar to the logistic NRC model.
Two of the collective inference methods show a significant difference from the rest.
On the one hand, iterative classification is significantly worse than the other.
It is most likely because it infers labels and not scores, which gives it less flexibility.
On the other hand, not combining RC with any CI proves more efficient than the alternative.
This is an interesting result since CIs have been shown to improve the performance of RC in other fields.
In the case of telcos, however, the churn signals might not be strong, common and interconnected enough for them to survive being spread through the network in multiple iterations, and the resulting scores are simply too diluted for the models to be of any use for making accurate predictions.
We also test the effect of building churn prediction models using non-relational classifiers, with and without scores from relational learners.
We show that the NRC performs better than the RL by themselves, but also that including the scores in the classifiers has a significant increase in their performance.
Thus the scores do help, when combined with other features in NRC.
However, when the performances of best performing relational learners are compared separately to those of the NRC models, there is not always a clear advantage of using the NRC models and, in this case, RLs do perform just as well, which indicates that they can be used on their own as churn prediction models.
Important consequences of the research questions can be identified for both business and scientific communities.
A business user, such as a retention manager in a telco, can use the results when creating a data driven campaign.
By using the result of this paper, a model with a non-relational classifier will ensure optimal results.
Because we have shown that the scores of relational learners  improve the model performance, enriching the dataset with scores from a relational classifier such as NLB is recommended. 
On the other hand, in terms of scientific contributions, our result that collective inference methods do not improve the performance of relational classifiers, as is the case in other fields \citep{jensen2004collective,sen2008collective}, is important.  
Furthermore, the conclusion that RL scores make a difference in the NRC models, suggests that "churn influence" is not only bound to a customer's first order neighborhood, but reaches further in the network. 
Finally, since many relational learners exist, our results provide a quantitative evaluation of these methods and how to apply them; which is an important contribution to the research domain of churn prediction in the telecommunication industry.

The performance of churn prediction models needs to be evaluated in order to compare different models and select the best one \citep{verbeke2012new}.
The choice of the performance measure should depend on the purpose of the model and take into account how it is to be used.  
In CCP, this means accurately identifying the customers with the highest propensity to churn, since the loss of these customers will result in the highest loss for the provider.
The commonly used AUC  and lift measures do not take into account the profit and costs which inevitably accompany a retention campaign, in contrast to the Maximum Profit Measure \citep{verbeke2012new,verbraken2013novel}.
This profit-centric measure evaluates models by optimizing the profit of the model and the fraction of the customers that should be offered promotions in order to get the highest profit.

\section{Conclusion}
\subsection{Main Contributions}
Social network analysis has been shown to make a difference when applied to customer churn prediction in telco.
From the definition of the network itself, to extracting insightful features, and building and evaluating the predictive models, multiple decisions need to be made at every step of the process.
In this study, we performed tests on some of these possibilities, investigated the impact of the construction of networks and the applied techniques on model performance and offered guidelines on how this can be done optimally.
Overall our study offers two main contributions to the existing literature.

Our first main contribution is an empirical comparison of the performance of relational learners and their combination with binary classifiers when predicting customer churn in telco.
The results were evaluated on a large number of CDR datasets, which allowed us to apply statistical tests to assess the significance of our results.
The CDR datasets originated from a number of telcos across the world and varied in both size and churn rate.
The performance of the resulting models was evaluated with four measures, which gave consistent and robust results.
A benchmarking of SNA methods for churn prediction in telco on this scale does not have precedence in the literature to the best of our knowledge.
An overview of our findings can be seen in Table \ref{T:RQA}.

The second main contribution of this paper is an exploratory study of various network architectures and how they effect model performance.
As far as we know, this has not been studied explicitly in the context of customer churn before.
Our results imply that network definition does matter for performance, with undirected, binary network resulting in the highest performance and recent connections having greater importance than older ones do.
Furthermore, the inclusion of contextual information that leads to an appropriate definition of the network is key in assuring maximum predictive capabilities.

\subsection{Future Work}
Although we have been able to provide answers to the research questions, many more still remain unanswered and can be viewed as objectives for future studies.

First of all, a vital component of each dataset was missing in our analyses: local variables.
These represent information about the customers as isolated entities without relations to other customers, such as demographic information, pricing plans, and financials, among others.
Previous studies have shown that models built with these variables alone perform worse than when network features are included as well \citep{verbeke2014social}.
We would like to see how this effect presents itself together with relational learners and include it in the experimental design of our study.

Secondly, only information about phone calls was used to build networks throughout the study.
It would be interesting to see what effect including networks created from text messages and even roaming information would have.

Thirdly, we have provided an insight into how network construction might effect model performance.
This was only done by the means of one relational learner and one dataset.
A more comprehensive study with more datasets and other types of models would offer the possibility to evaluate this effect statistically and, as a result, provide guidelines on how to construct the network to achieve the highest performance.
In addition, since we only used one classifier, we were unable to verify whether there is correlation between network construction and classifier when it comes to model performance.
Finally, an in-depth study would allow us to determine the optimal way of constructing networks, with regard to edges and weights, and to determine the actual timespan of CDR data that are needed to achieve best performance.

\scriptsize{
\bibliography{benchmark_bib}

\begin{thebibliography}{64}
\expandafter\ifx\csname natexlab\endcsname\relax\def\natexlab#1{#1}\fi
\providecommand{\url}[1]{\texttt{#1}}
\providecommand{\href}[2]{#2}
\providecommand{\path}[1]{#1}
\providecommand{\DOIprefix}{doi:}
\providecommand{\ArXivprefix}{arXiv:}
\providecommand{\URLprefix}{URL: }
\providecommand{\Pubmedprefix}{pmid:}
\providecommand{\doi}[1]{\href{http://dx.doi.org/#1}{\path{#1}}}
\providecommand{\Pubmed}[1]{\href{pmid:#1}{\path{#1}}}
\providecommand{\bibinfo}[2]{#2}
\ifx\xfnm\relax \def\xfnm[#1]{\unskip,\space#1}\fi
\bibitem[{Ahn et~al.(2006)Ahn, Han \& Lee}]{ahn2006customer}
\bibinfo{author}{Ahn, J.-H.}, \bibinfo{author}{Han, S.-P.}, \&
  \bibinfo{author}{Lee, Y.-S.} (\bibinfo{year}{2006}).
\newblock \bibinfo{title}{Customer churn analysis: Churn determinants and
  mediation effects of partial defection in the korean mobile
  telecommunications service industry}.
\newblock {\it \bibinfo{journal}{Telecommunications policy}\/},  {\it
  \bibinfo{volume}{30}\/}, \bibinfo{pages}{552--568}.
\bibitem[{Ali \& Ar{\i}t{\"u}rk(2014)}]{ali2014dynamic}
\bibinfo{author}{Ali, {\"O}.~G.}, \& \bibinfo{author}{Ar{\i}t{\"u}rk, U.}
  (\bibinfo{year}{2014}).
\newblock \bibinfo{title}{Dynamic churn prediction framework with more
  effective use of rare event data: The case of private banking}.
\newblock {\it \bibinfo{journal}{Expert Systems with Applications}\/},  {\it
  \bibinfo{volume}{41}\/}, \bibinfo{pages}{7889--7903}.
\bibitem[{Anil~Kumar \& Ravi(2008)}]{anil2008predicting}
\bibinfo{author}{Anil~Kumar, D.}, \& \bibinfo{author}{Ravi, V.}
  (\bibinfo{year}{2008}).
\newblock \bibinfo{title}{Predicting credit card customer churn in banks using
  data mining}.
\newblock {\it \bibinfo{journal}{International Journal of Data Analysis
  Techniques and Strategies}\/},  {\it \bibinfo{volume}{1}\/},
  \bibinfo{pages}{4--28}.
\bibitem[{Athanassopoulos(2000)}]{athanassopoulos2000customer}
\bibinfo{author}{Athanassopoulos, A.~D.} (\bibinfo{year}{2000}).
\newblock \bibinfo{title}{Customer satisfaction cues to support market
  segmentation and explain switching behavior}.
\newblock {\it \bibinfo{journal}{Journal of business research}\/},  {\it
  \bibinfo{volume}{47}\/}, \bibinfo{pages}{191--207}.
\bibitem[{Backiel et~al.(2014)Backiel, Baesens \&
  Claeskens}]{backiel2014mining}
\bibinfo{author}{Backiel, A.}, \bibinfo{author}{Baesens, B.}, \&
  \bibinfo{author}{Claeskens, G.} (\bibinfo{year}{2014}).
\newblock \bibinfo{title}{Mining telecommunication networks to enhance customer
  lifetime predictions}.
\newblock In {\it \bibinfo{booktitle}{Artificial Intelligence and Soft
  Computing}\/} (pp. \bibinfo{pages}{15--26}).
\newblock \bibinfo{organization}{Springer}.
\bibitem[{Backiel et~al.(2015)Backiel, Verbinnen, Baesens \&
  Claeskens}]{backiel2015combining}
\bibinfo{author}{Backiel, A.}, \bibinfo{author}{Verbinnen, Y.},
  \bibinfo{author}{Baesens, B.}, \& \bibinfo{author}{Claeskens, G.}
  (\bibinfo{year}{2015}).
\newblock \bibinfo{title}{Combining local and social network classifiers to
  improve churn prediction}.
\newblock In {\it \bibinfo{booktitle}{Proceedings of the 2015 IEEE/ACM
  International Conference on Advances in Social Networks Analysis and Mining
  2015}\/} (pp. \bibinfo{pages}{651--658}).
\newblock \bibinfo{organization}{ACM}.
\bibitem[{Baesens et~al.(2015)Baesens, Van~Vlasselaer \&
  Verbeke}]{baesens2015fraud}
\bibinfo{author}{Baesens, B.}, \bibinfo{author}{Van~Vlasselaer, V.}, \&
  \bibinfo{author}{Verbeke, W.} (\bibinfo{year}{2015}).
\newblock {\it \bibinfo{title}{Fraud Analytics Using Descriptive, Predictive,
  and Social Network Techniques: A Guide to Data Science for Fraud
  Detection}\/}.
\newblock Wiley and SAS Business Series.
\newblock \bibinfo{publisher}{Wiley}.
\bibitem[{Baras et~al.(2014)Baras, Ronen \& Yom-Tov}]{baras2014effect}
\bibinfo{author}{Baras, D.}, \bibinfo{author}{Ronen, A.}, \&
  \bibinfo{author}{Yom-Tov, E.} (\bibinfo{year}{2014}).
\newblock \bibinfo{title}{The effect of social affinity and predictive horizon
  on churn prediction using diffusion modeling}.
\newblock {\it \bibinfo{journal}{Social Network Analysis and Mining}\/},  {\it
  \bibinfo{volume}{4}\/}, \bibinfo{pages}{1--12}.
\bibitem[{Benedek et~al.(2014)Benedek, Lubl{\'o}y \&
  Vastag}]{benedek2014importance}
\bibinfo{author}{Benedek, G.}, \bibinfo{author}{Lubl{\'o}y, {\'A}.}, \&
  \bibinfo{author}{Vastag, G.} (\bibinfo{year}{2014}).
\newblock \bibinfo{title}{The importance of social embeddedness: Churn models
  at mobile providers}.
\newblock {\it \bibinfo{journal}{Decision Sciences}\/},  {\it
  \bibinfo{volume}{45}\/}, \bibinfo{pages}{175--201}.
\bibitem[{Berson \& Smith(2002)}]{berson2002building}
\bibinfo{author}{Berson, A.}, \& \bibinfo{author}{Smith, S.~J.}
  (\bibinfo{year}{2002}).
\newblock {\it \bibinfo{title}{Building data mining applications for CRM}\/}.
\newblock \bibinfo{publisher}{McGraw-Hill, Inc.}
\bibitem[{Chakrabarti et~al.(1998)Chakrabarti, Dom \&
  Indyk}]{chakrabarti1998enhanced}
\bibinfo{author}{Chakrabarti, S.}, \bibinfo{author}{Dom, B.}, \&
  \bibinfo{author}{Indyk, P.} (\bibinfo{year}{1998}).
\newblock \bibinfo{title}{Enhanced hypertext categorization using hyperlinks}.
\newblock In {\it \bibinfo{booktitle}{ACM SIGMOD Record}\/} (pp.
  \bibinfo{pages}{307--318}).
\newblock \bibinfo{organization}{ACM} volume~\bibinfo{volume}{27}.
\bibitem[{Chen et~al.(2012)Chen, Fan \& Sun}]{chen2012hierarchical}
\bibinfo{author}{Chen, Z.-Y.}, \bibinfo{author}{Fan, Z.-P.}, \&
  \bibinfo{author}{Sun, M.} (\bibinfo{year}{2012}).
\newblock \bibinfo{title}{A hierarchical multiple kernel support vector machine
  for customer churn prediction using longitudinal behavioral data}.
\newblock {\it \bibinfo{journal}{European Journal of operational research}\/},
  {\it \bibinfo{volume}{223}\/}, \bibinfo{pages}{461--472}.
\bibitem[{Coussement \& Van~den Poel(2008)}]{coussement2008churn}
\bibinfo{author}{Coussement, K.}, \& \bibinfo{author}{Van~den Poel, D.}
  (\bibinfo{year}{2008}).
\newblock \bibinfo{title}{Churn prediction in subscription services: An
  application of support vector machines while comparing two
  parameter-selection techniques}.
\newblock {\it \bibinfo{journal}{Expert systems with applications}\/},  {\it
  \bibinfo{volume}{34}\/}, \bibinfo{pages}{313--327}.
\bibitem[{Dasgupta et~al.(2008)Dasgupta, Singh, Viswanathan, Chakraborty,
  Mukherjea, Nanavati \& Joshi}]{dasgupta2008social}
\bibinfo{author}{Dasgupta, K.}, \bibinfo{author}{Singh, R.},
  \bibinfo{author}{Viswanathan, B.}, \bibinfo{author}{Chakraborty, D.},
  \bibinfo{author}{Mukherjea, S.}, \bibinfo{author}{Nanavati, A.~A.}, \&
  \bibinfo{author}{Joshi, A.} (\bibinfo{year}{2008}).
\newblock \bibinfo{title}{Social ties and their relevance to churn in mobile
  telecom networks}.
\newblock In {\it \bibinfo{booktitle}{Proceedings of the 11th international
  conference on Extending database technology: Advances in database
  technology}\/} (pp. \bibinfo{pages}{668--677}).
\newblock \bibinfo{organization}{ACM}.
\bibitem[{De~Bock \& Van~den Poel(2011)}]{de2011empirical}
\bibinfo{author}{De~Bock, K.~W.}, \& \bibinfo{author}{Van~den Poel, D.}
  (\bibinfo{year}{2011}).
\newblock \bibinfo{title}{An empirical evaluation of rotation-based ensemble
  classifiers for customer churn prediction}.
\newblock {\it \bibinfo{journal}{Expert Systems with Applications}\/},  {\it
  \bibinfo{volume}{38}\/}, \bibinfo{pages}{12293--12301}.
\bibitem[{Demsar(2006)}]{demvsar2006statistical}
\bibinfo{author}{Demsar, J.} (\bibinfo{year}{2006}).
\newblock \bibinfo{title}{Statistical comparisons of classifiers over multiple
  data sets}.
\newblock {\it \bibinfo{journal}{The Journal of Machine Learning Research}\/},
  {\it \bibinfo{volume}{7}\/}, \bibinfo{pages}{1--30}.
\bibitem[{Dierkes et~al.(2011)Dierkes, Bichler \&
  Krishnan}]{dierkes2011estimating}
\bibinfo{author}{Dierkes, T.}, \bibinfo{author}{Bichler, M.}, \&
  \bibinfo{author}{Krishnan, R.} (\bibinfo{year}{2011}).
\newblock \bibinfo{title}{Estimating the effect of word of mouth on churn and
  cross-buying in the mobile phone market with markov logic networks}.
\newblock {\it \bibinfo{journal}{Decision Support Systems}\/},  {\it
  \bibinfo{volume}{51}\/}, \bibinfo{pages}{361--371}.
\bibitem[{Fawcett(2006)}]{fawcett2006introduction}
\bibinfo{author}{Fawcett, T.} (\bibinfo{year}{2006}).
\newblock \bibinfo{title}{An introduction to roc analysis}.
\newblock {\it \bibinfo{journal}{Pattern recognition letters}\/},  {\it
  \bibinfo{volume}{27}\/}, \bibinfo{pages}{861--874}.
\bibitem[{Ganesh et~al.(2000)Ganesh, Arnold \&
  Reynolds}]{ganesh2000understanding}
\bibinfo{author}{Ganesh, J.}, \bibinfo{author}{Arnold, M.~J.}, \&
  \bibinfo{author}{Reynolds, K.~E.} (\bibinfo{year}{2000}).
\newblock \bibinfo{title}{Understanding the customer base of service providers:
  an examination of the differences between switchers and stayers}.
\newblock {\it \bibinfo{journal}{Journal of marketing}\/},  {\it
  \bibinfo{volume}{64}\/}, \bibinfo{pages}{65--87}.
\bibitem[{Geman \& Geman(1984)}]{geman1984stochastic}
\bibinfo{author}{Geman, S.}, \& \bibinfo{author}{Geman, D.}
  (\bibinfo{year}{1984}).
\newblock \bibinfo{title}{Stochastic relaxation, gibbs distributions, and the
  bayesian restoration of images}.
\newblock {\it \bibinfo{journal}{Pattern Analysis and Machine Intelligence,
  IEEE Transactions on}\/},  (pp. \bibinfo{pages}{721--741}).
\bibitem[{Glady et~al.(2009)Glady, Baesens \& Croux}]{glady2009modeling}
\bibinfo{author}{Glady, N.}, \bibinfo{author}{Baesens, B.}, \&
  \bibinfo{author}{Croux, C.} (\bibinfo{year}{2009}).
\newblock \bibinfo{title}{Modeling churn using customer lifetime value}.
\newblock {\it \bibinfo{journal}{European Journal of Operational Research}\/},
  {\it \bibinfo{volume}{197}\/}, \bibinfo{pages}{402--411}.
\bibitem[{Guill{\'e}n et~al.(2012)Guill{\'e}n, Nielsen, Scheike \&
  P{\'e}rez-Mar{\'\i}n}]{guillen2012time}
\bibinfo{author}{Guill{\'e}n, M.}, \bibinfo{author}{Nielsen, J.~P.},
  \bibinfo{author}{Scheike, T.~H.}, \& \bibinfo{author}{P{\'e}rez-Mar{\'\i}n,
  A.~M.} (\bibinfo{year}{2012}).
\newblock \bibinfo{title}{Time-varying effects in the analysis of customer
  loyalty: A case study in insurance}.
\newblock {\it \bibinfo{journal}{Expert Systems with Applications}\/},  {\it
  \bibinfo{volume}{39}\/}, \bibinfo{pages}{3551--3558}.
\bibitem[{G{\"u}nther et~al.(2014)G{\"u}nther, Tvete, Aas, Sandnes \&
  Borgan}]{gunther2014modelling}
\bibinfo{author}{G{\"u}nther, C.-C.}, \bibinfo{author}{Tvete, I.~F.},
  \bibinfo{author}{Aas, K.}, \bibinfo{author}{Sandnes, G.~I.}, \&
  \bibinfo{author}{Borgan, {\O}.} (\bibinfo{year}{2014}).
\newblock \bibinfo{title}{Modelling and predicting customer churn from an
  insurance company}.
\newblock {\it \bibinfo{journal}{Scandinavian Actuarial Journal}\/},  {\it
  \bibinfo{volume}{2014}\/}, \bibinfo{pages}{58--71}.
\bibitem[{Haenlein(2013)}]{haenlein2013social}
\bibinfo{author}{Haenlein, M.} (\bibinfo{year}{2013}).
\newblock \bibinfo{title}{Social interactions in customer churn decisions: The
  impact of relationship directionality}.
\newblock {\it \bibinfo{journal}{International Journal of Research in
  Marketing}\/},  {\it \bibinfo{volume}{30}\/}, \bibinfo{pages}{236--248}.
\bibitem[{Hand(2009)}]{hand2009measuring}
\bibinfo{author}{Hand, D.~J.} (\bibinfo{year}{2009}).
\newblock \bibinfo{title}{Measuring classifier performance: a coherent
  alternative to the area under the roc curve}.
\newblock {\it \bibinfo{journal}{Machine learning}\/},  {\it
  \bibinfo{volume}{77}\/}, \bibinfo{pages}{103--123}.
\bibitem[{Huang et~al.(2012)Huang, Kechadi \& Buckley}]{huang2012customer}
\bibinfo{author}{Huang, B.}, \bibinfo{author}{Kechadi, M.~T.}, \&
  \bibinfo{author}{Buckley, B.} (\bibinfo{year}{2012}).
\newblock \bibinfo{title}{Customer churn prediction in telecommunications}.
\newblock {\it \bibinfo{journal}{Expert Systems with Applications}\/},  {\it
  \bibinfo{volume}{39}\/}, \bibinfo{pages}{1414--1425}.
\bibitem[{Hung et~al.(2006)Hung, Yen \& Wang}]{hung2006applying}
\bibinfo{author}{Hung, S.-Y.}, \bibinfo{author}{Yen, D.~C.}, \&
  \bibinfo{author}{Wang, H.-Y.} (\bibinfo{year}{2006}).
\newblock \bibinfo{title}{Applying data mining to telecom churn management}.
\newblock {\it \bibinfo{journal}{Expert Systems with Applications}\/},  {\it
  \bibinfo{volume}{31}\/}, \bibinfo{pages}{515--524}.
\bibitem[{Jensen et~al.(2004)Jensen, Neville \&
  Gallagher}]{jensen2004collective}
\bibinfo{author}{Jensen, D.}, \bibinfo{author}{Neville, J.}, \&
  \bibinfo{author}{Gallagher, B.} (\bibinfo{year}{2004}).
\newblock \bibinfo{title}{Why collective inference improves relational
  classification}.
\newblock In {\it \bibinfo{booktitle}{Proceedings of the tenth ACM SIGKDD
  international conference on Knowledge discovery and data mining}\/} (pp.
  \bibinfo{pages}{593--598}).
\newblock \bibinfo{organization}{ACM}.
\bibitem[{Khan et~al.(2010)Khan, Jamwal \& Sepehri}]{khan2010applying}
\bibinfo{author}{Khan, A.~A.}, \bibinfo{author}{Jamwal, S.}, \&
  \bibinfo{author}{Sepehri, M.} (\bibinfo{year}{2010}).
\newblock \bibinfo{title}{Applying data mining to customer churn prediction in
  an internet service provider}.
\newblock {\it \bibinfo{journal}{International Journal of Computer
  Applications}\/},  {\it \bibinfo{volume}{9}\/}, \bibinfo{pages}{8--14}.
\bibitem[{Kim et~al.(2014)Kim, Jun \& Lee}]{kim2014improved}
\bibinfo{author}{Kim, K.}, \bibinfo{author}{Jun, C.-H.}, \&
  \bibinfo{author}{Lee, J.} (\bibinfo{year}{2014}).
\newblock \bibinfo{title}{Improved churn prediction in telecommunication
  industry by analyzing a large network}.
\newblock {\it \bibinfo{journal}{Expert Systems with Applications}\/},  {\it
  \bibinfo{volume}{41}\/}, \bibinfo{pages}{6575--6584}.
\bibitem[{Kusuma et~al.(2013)Kusuma, Radosavljevik, Takes \& van~der
  Putten}]{kusuma2013combining}
\bibinfo{author}{Kusuma, P.~D.}, \bibinfo{author}{Radosavljevik, D.},
  \bibinfo{author}{Takes, F.~W.}, \& \bibinfo{author}{van~der Putten, P.}
  (\bibinfo{year}{2013}).
\newblock \bibinfo{title}{Combining customer attribute and social network
  mining for prepaid mobile churn prediction}.
\newblock In {\it \bibinfo{booktitle}{Proc. the 23rd Annual Belgian Dutch
  Conference on Machine Learning (BENELEARN)}\/} (pp. \bibinfo{pages}{50--58}).
\bibitem[{Larivi{\`e}re \& Van~den Poel(2004)}]{lariviere2004investigating}
\bibinfo{author}{Larivi{\`e}re, B.}, \& \bibinfo{author}{Van~den Poel, D.}
  (\bibinfo{year}{2004}).
\newblock \bibinfo{title}{Investigating the role of product features in
  preventing customer churn, by using survival analysis and choice modeling:
  The case of financial services}.
\newblock {\it \bibinfo{journal}{Expert Systems with Applications}\/},  {\it
  \bibinfo{volume}{27}\/}, \bibinfo{pages}{277--285}.
\bibitem[{Lessmann et~al.(2015)Lessmann, Baesens, Seow \&
  Thomas}]{lessmann2015benchmarking}
\bibinfo{author}{Lessmann, S.}, \bibinfo{author}{Baesens, B.},
  \bibinfo{author}{Seow, H.-V.}, \& \bibinfo{author}{Thomas, L.~C.}
  (\bibinfo{year}{2015}).
\newblock \bibinfo{title}{Benchmarking state-of-the-art classification
  algorithms for credit scoring: An update of research}.
\newblock {\it \bibinfo{journal}{European Journal of Operational Research}\/},
  {\it \bibinfo{volume}{247}\/}, \bibinfo{pages}{124--136}.
\bibitem[{Lu \& Getoor(2003)}]{lu2003link}
\bibinfo{author}{Lu, Q.}, \& \bibinfo{author}{Getoor, L.}
  (\bibinfo{year}{2003}).
\newblock \bibinfo{title}{Link-based classification}.
\newblock In {\it \bibinfo{booktitle}{ICML}\/} (pp. \bibinfo{pages}{496--503}).
\newblock volume~\bibinfo{volume}{3}.
\bibitem[{Macskassy \& Provost(2007)}]{macskassy2007classification}
\bibinfo{author}{Macskassy, S.~A.}, \& \bibinfo{author}{Provost, F.}
  (\bibinfo{year}{2007}).
\newblock \bibinfo{title}{Classification in networked data: A toolkit and a
  univariate case study}.
\newblock {\it \bibinfo{journal}{The Journal of Machine Learning Research}\/},
  {\it \bibinfo{volume}{8}\/}, \bibinfo{pages}{935--983}.
\bibitem[{Miritello et~al.(2013)Miritello, Moro, Lara, Mart{\'\i}nez-L{\'o}pez,
  Belchamber, Roberts \& Dunbar}]{miritello2013time}
\bibinfo{author}{Miritello, G.}, \bibinfo{author}{Moro, E.},
  \bibinfo{author}{Lara, R.}, \bibinfo{author}{Mart{\'\i}nez-L{\'o}pez, R.},
  \bibinfo{author}{Belchamber, J.}, \bibinfo{author}{Roberts, S.~G.}, \&
  \bibinfo{author}{Dunbar, R.~I.} (\bibinfo{year}{2013}).
\newblock \bibinfo{title}{Time as a limited resource: Communication strategy in
  mobile phone networks}.
\newblock {\it \bibinfo{journal}{Social Networks}\/},  {\it
  \bibinfo{volume}{35}\/}, \bibinfo{pages}{89--95}.
\bibitem[{Modani et~al.(2013)Modani, Dey, Gupta \& Godbole}]{modani2013cdr}
\bibinfo{author}{Modani, N.}, \bibinfo{author}{Dey, K.},
  \bibinfo{author}{Gupta, R.}, \& \bibinfo{author}{Godbole, S.}
  (\bibinfo{year}{2013}).
\newblock \bibinfo{title}{Cdr analysis based telco churn prediction and
  customer behavior insights: A case study}.
\newblock In {\it \bibinfo{booktitle}{Web Information Systems Engineering--WISE
  2013}\/} (pp. \bibinfo{pages}{256--269}).
\newblock \bibinfo{publisher}{Springer}.
\bibitem[{Mozer et~al.(2000)Mozer, Wolniewicz, Grimes, Johnson \&
  Kaushansky}]{mozer2000predicting}
\bibinfo{author}{Mozer, M.~C.}, \bibinfo{author}{Wolniewicz, R.},
  \bibinfo{author}{Grimes, D.~B.}, \bibinfo{author}{Johnson, E.}, \&
  \bibinfo{author}{Kaushansky, H.} (\bibinfo{year}{2000}).
\newblock \bibinfo{title}{Predicting subscriber dissatisfaction and improving
  retention in the wireless telecommunications industry}.
\newblock {\it \bibinfo{journal}{Neural Networks, IEEE Transactions on}\/},
  {\it \bibinfo{volume}{11}\/}, \bibinfo{pages}{690--696}.
\bibitem[{Naboulsi et~al.(2015)Naboulsi, Fiore, Ribot \&
  Stanica}]{naboulsi2015large}
\bibinfo{author}{Naboulsi, D.}, \bibinfo{author}{Fiore, M.},
  \bibinfo{author}{Ribot, S.}, \& \bibinfo{author}{Stanica, R.}
  (\bibinfo{year}{2015}).
\newblock \bibinfo{title}{Large-scale mobile traffic analysis: a survey}.
\newblock {\it \bibinfo{journal}{IEEE Communications Surveys \& Tutorials}\/},
  {\it \bibinfo{volume}{18}\/}, \bibinfo{pages}{124--161}.
\bibitem[{Nanavati et~al.(2008)Nanavati, Singh, Chakraborty, Dasgupta,
  Mukherjea, Das, Gurumurthy \& Joshi}]{nanavati2008analyzing}
\bibinfo{author}{Nanavati, A.~A.}, \bibinfo{author}{Singh, R.},
  \bibinfo{author}{Chakraborty, D.}, \bibinfo{author}{Dasgupta, K.},
  \bibinfo{author}{Mukherjea, S.}, \bibinfo{author}{Das, G.},
  \bibinfo{author}{Gurumurthy, S.}, \& \bibinfo{author}{Joshi, A.}
  (\bibinfo{year}{2008}).
\newblock \bibinfo{title}{Analyzing the structure and evolution of massive
  telecom graphs}.
\newblock {\it \bibinfo{journal}{IEEE Transactions on Knowledge and Data
  Engineering}\/},  {\it \bibinfo{volume}{20}\/}, \bibinfo{pages}{703--718}.
\bibitem[{Neslin et~al.(2006)Neslin, Gupta, Kamakura, Lu \&
  Mason}]{neslin2006defection}
\bibinfo{author}{Neslin, S.~A.}, \bibinfo{author}{Gupta, S.},
  \bibinfo{author}{Kamakura, W.}, \bibinfo{author}{Lu, J.}, \&
  \bibinfo{author}{Mason, C.~H.} (\bibinfo{year}{2006}).
\newblock \bibinfo{title}{Defection detection: Measuring and understanding the
  predictive accuracy of customer churn models}.
\newblock {\it \bibinfo{journal}{Journal of marketing research}\/},  {\it
  \bibinfo{volume}{43}\/}, \bibinfo{pages}{204--211}.
\bibitem[{Newman(2010)}]{newman2010networks}
\bibinfo{author}{Newman, M.} (\bibinfo{year}{2010}).
\newblock {\it \bibinfo{title}{Networks: an introduction}\/}.
\newblock \bibinfo{publisher}{OUP Oxford}.
\bibitem[{Ngonmang et~al.(2012)Ngonmang, Viennet \&
  Tchuente}]{ngonmang2012churn}
\bibinfo{author}{Ngonmang, B.}, \bibinfo{author}{Viennet, E.}, \&
  \bibinfo{author}{Tchuente, M.} (\bibinfo{year}{2012}).
\newblock \bibinfo{title}{Churn prediction in a real online social network
  using local community analysis}.
\newblock In {\it \bibinfo{booktitle}{Proceedings of the 2012 International
  Conference on Advances in Social Networks Analysis and Mining (ASONAM
  2012)}\/} (pp. \bibinfo{pages}{282--288}).
\newblock \bibinfo{organization}{IEEE Computer Society}.
\bibitem[{Park et~al.(2012)Park, Huh, Oh \& Han}]{park2012social}
\bibinfo{author}{Park, S.-H.}, \bibinfo{author}{Huh, S.-Y.},
  \bibinfo{author}{Oh, W.}, \& \bibinfo{author}{Han, S.~P.}
  (\bibinfo{year}{2012}).
\newblock \bibinfo{title}{A social network-based inference model for validating
  customer profile data}.
\newblock {\it \bibinfo{journal}{MIS Quarterly}\/},  {\it
  \bibinfo{volume}{36}\/}, \bibinfo{pages}{1217--1237}.
\bibitem[{Phadke et~al.(2013)Phadke, Uzunalioglu, Mendiratta, Kushnir \&
  Doran}]{phadke2013prediction}
\bibinfo{author}{Phadke, C.}, \bibinfo{author}{Uzunalioglu, H.},
  \bibinfo{author}{Mendiratta, V.~B.}, \bibinfo{author}{Kushnir, D.}, \&
  \bibinfo{author}{Doran, D.} (\bibinfo{year}{2013}).
\newblock \bibinfo{title}{Prediction of subscriber churn using social network
  analysis}.
\newblock {\it \bibinfo{journal}{Bell Labs Technical Journal}\/},  {\it
  \bibinfo{volume}{17}\/}, \bibinfo{pages}{63--75}.
\bibitem[{Van~den Poel \& Lariviere(2004)}]{van2004customer}
\bibinfo{author}{Van~den Poel, D.}, \& \bibinfo{author}{Lariviere, B.}
  (\bibinfo{year}{2004}).
\newblock \bibinfo{title}{Customer attrition analysis for financial services
  using proportional hazard models}.
\newblock {\it \bibinfo{journal}{European journal of operational research}\/},
  {\it \bibinfo{volume}{157}\/}, \bibinfo{pages}{196--217}.
\bibitem[{Raeder et~al.(2011)Raeder, Lizardo, Hachen \&
  Chawla}]{raeder2011predictors}
\bibinfo{author}{Raeder, T.}, \bibinfo{author}{Lizardo, O.},
  \bibinfo{author}{Hachen, D.}, \& \bibinfo{author}{Chawla, N.~V.}
  (\bibinfo{year}{2011}).
\newblock \bibinfo{title}{Predictors of short-term decay of cell phone contacts
  in a large scale communication network}.
\newblock {\it \bibinfo{journal}{Social Networks}\/},  {\it
  \bibinfo{volume}{33}\/}, \bibinfo{pages}{245--257}.
\bibitem[{Rehman \& Raza~Ali(2015)}]{rehman2015customer}
\bibinfo{author}{Rehman, A.}, \& \bibinfo{author}{Raza~Ali, A.}
  (\bibinfo{year}{2015}).
\newblock \bibinfo{title}{Customer churn prediction, segmentation and fraud
  detection in telecommunication industry}, .
\bibitem[{Richter et~al.(2010)Richter, Yom-Tov \&
  Slonim}]{richter2010predicting}
\bibinfo{author}{Richter, Y.}, \bibinfo{author}{Yom-Tov, E.}, \&
  \bibinfo{author}{Slonim, N.} (\bibinfo{year}{2010}).
\newblock \bibinfo{title}{Predicting customer churn in mobile networks through
  analysis of social groups.}
\newblock In {\it \bibinfo{booktitle}{SDM}\/} (pp. \bibinfo{pages}{732--741}).
\newblock volume \bibinfo{volume}{2010}.
\bibitem[{Rocchio(1971)}]{rocchio1971relevance}
\bibinfo{author}{Rocchio, J.~J.} (\bibinfo{year}{1971}).
\newblock \bibinfo{title}{Relevance feedback in information retrieval}, .
\bibitem[{Sarraute et~al.(2015)Sarraute, Brea, Burroni \&
  Blanc}]{sarraute2015inference}
\bibinfo{author}{Sarraute, C.}, \bibinfo{author}{Brea, J.},
  \bibinfo{author}{Burroni, J.}, \& \bibinfo{author}{Blanc, P.}
  (\bibinfo{year}{2015}).
\newblock \bibinfo{title}{Inference of demographic attributes based on mobile
  phone usage patterns and social network topology}.
\newblock {\it \bibinfo{journal}{Social Network Analysis and Mining}\/},  {\it
  \bibinfo{volume}{5}\/}, \bibinfo{pages}{1--18}.
\bibitem[{Sen et~al.(2008)Sen, Namata, Bilgic, Getoor, Galligher \&
  Eliassi-Rad}]{sen2008collective}
\bibinfo{author}{Sen, P.}, \bibinfo{author}{Namata, G.},
  \bibinfo{author}{Bilgic, M.}, \bibinfo{author}{Getoor, L.},
  \bibinfo{author}{Galligher, B.}, \& \bibinfo{author}{Eliassi-Rad, T.}
  (\bibinfo{year}{2008}).
\newblock \bibinfo{title}{Collective classification in network data}.
\newblock {\it \bibinfo{journal}{AI magazine}\/},  {\it
  \bibinfo{volume}{29}\/}, \bibinfo{pages}{93}.
\bibitem[{Tomar et~al.(2010)Tomar, Asnani, Karandikar, Chander, Agrawal \&
  Kapadia}]{tomar2010social}
\bibinfo{author}{Tomar, V.}, \bibinfo{author}{Asnani, H.},
  \bibinfo{author}{Karandikar, A.}, \bibinfo{author}{Chander, V.},
  \bibinfo{author}{Agrawal, S.}, \& \bibinfo{author}{Kapadia, P.}
  (\bibinfo{year}{2010}).
\newblock \bibinfo{title}{Social network analysis of the short message
  service}.
\newblock In {\it \bibinfo{booktitle}{2010 National Conference on
  Communications, Chennai}\/} (pp. \bibinfo{pages}{1--5}).
\newblock \bibinfo{organization}{Citeseer}.
\bibitem[{Tuff{\'e}ry(2011)}]{tuffery2011data}
\bibinfo{author}{Tuff{\'e}ry, S.} (\bibinfo{year}{2011}).
\newblock {\it \bibinfo{title}{Data mining and statistics for decision
  making}\/}.
\newblock \bibinfo{publisher}{John Wiley \& Sons}.
\bibitem[{Vafeiadis et~al.(2015)Vafeiadis, Diamantaras, Sarigiannidis \&
  Chatzisavvas}]{vafeiadis2015comparison}
\bibinfo{author}{Vafeiadis, T.}, \bibinfo{author}{Diamantaras, K.~I.},
  \bibinfo{author}{Sarigiannidis, G.}, \& \bibinfo{author}{Chatzisavvas, K.~C.}
  (\bibinfo{year}{2015}).
\newblock \bibinfo{title}{A comparison of machine learning techniques for
  customer churn prediction}.
\newblock {\it \bibinfo{journal}{Simulation Modelling Practice and Theory}\/},
  {\it \bibinfo{volume}{55}\/}, \bibinfo{pages}{1--9}.
\bibitem[{Van~Vlasselaer et~al.(2015)Van~Vlasselaer, Bravo, Caelen,
  Eliassi-Rad, Akoglu, Snoeck \& Baesens}]{van2015apate}
\bibinfo{author}{Van~Vlasselaer, V.}, \bibinfo{author}{Bravo, C.},
  \bibinfo{author}{Caelen, O.}, \bibinfo{author}{Eliassi-Rad, T.},
  \bibinfo{author}{Akoglu, L.}, \bibinfo{author}{Snoeck, M.}, \&
  \bibinfo{author}{Baesens, B.} (\bibinfo{year}{2015}).
\newblock \bibinfo{title}{Apate: A novel approach for automated credit card
  transaction fraud detection using network-based extensions}.
\newblock {\it \bibinfo{journal}{Decision Support Systems}\/},  {\it
  \bibinfo{volume}{75}\/}, \bibinfo{pages}{38--48}.
\bibitem[{Van~Vlasselaer et~al.(2016)Van~Vlasselaer, Eliassi-Rad, Akoglu,
  Snoeck \& Baesens}]{van2014gotcha}
\bibinfo{author}{Van~Vlasselaer, V.}, \bibinfo{author}{Eliassi-Rad, T.},
  \bibinfo{author}{Akoglu, L.}, \bibinfo{author}{Snoeck, M.}, \&
  \bibinfo{author}{Baesens, B.} (\bibinfo{year}{2016}).
\newblock \bibinfo{title}{Gotcha! network-based fraud detection for social
  security fraud}.
\newblock {\it \bibinfo{journal}{Management Science, accepted}\/}, .
\bibitem[{Verbeke et~al.(2012)Verbeke, Dejaeger, Martens, Hur \&
  Baesens}]{verbeke2012new}
\bibinfo{author}{Verbeke, W.}, \bibinfo{author}{Dejaeger, K.},
  \bibinfo{author}{Martens, D.}, \bibinfo{author}{Hur, J.}, \&
  \bibinfo{author}{Baesens, B.} (\bibinfo{year}{2012}).
\newblock \bibinfo{title}{New insights into churn prediction in the
  telecommunication sector: A profit driven data mining approach}.
\newblock {\it \bibinfo{journal}{European Journal of Operational Research}\/},
  {\it \bibinfo{volume}{218}\/}, \bibinfo{pages}{211--229}.
\bibitem[{Verbeke et~al.(2014)Verbeke, Martens \& Baesens}]{verbeke2014social}
\bibinfo{author}{Verbeke, W.}, \bibinfo{author}{Martens, D.}, \&
  \bibinfo{author}{Baesens, B.} (\bibinfo{year}{2014}).
\newblock \bibinfo{title}{Social network analysis for customer churn
  prediction}.
\newblock {\it \bibinfo{journal}{Applied Soft Computing}\/},  {\it
  \bibinfo{volume}{14}\/}, \bibinfo{pages}{431--446}.
\bibitem[{Verbraken et~al.(2014)Verbraken, Bravo, Weber \&
  Baesens}]{verbraken2014development}
\bibinfo{author}{Verbraken, T.}, \bibinfo{author}{Bravo, C.},
  \bibinfo{author}{Weber, R.}, \& \bibinfo{author}{Baesens, B.}
  (\bibinfo{year}{2014}).
\newblock \bibinfo{title}{Development and application of consumer credit
  scoring models using profit-based classification measures}.
\newblock {\it \bibinfo{journal}{European Journal of Operational Research}\/},
  {\it \bibinfo{volume}{238}\/}, \bibinfo{pages}{505--513}.
\bibitem[{Verbraken et~al.(2013)Verbraken, Verbeke \&
  Baesens}]{verbraken2013novel}
\bibinfo{author}{Verbraken, T.}, \bibinfo{author}{Verbeke, W.}, \&
  \bibinfo{author}{Baesens, B.} (\bibinfo{year}{2013}).
\newblock \bibinfo{title}{A novel profit maximizing metric for measuring
  classification performance of customer churn prediction models}.
\newblock {\it \bibinfo{journal}{Knowledge and Data Engineering, IEEE
  Transactions on}\/},  {\it \bibinfo{volume}{25}\/},
  \bibinfo{pages}{961--973}.
\bibitem[{Xie et~al.(2009)Xie, Li, Ngai \& Ying}]{xie2009customer}
\bibinfo{author}{Xie, Y.}, \bibinfo{author}{Li, X.}, \bibinfo{author}{Ngai,
  E.}, \& \bibinfo{author}{Ying, W.} (\bibinfo{year}{2009}).
\newblock \bibinfo{title}{Customer churn prediction using improved balanced
  random forests}.
\newblock {\it \bibinfo{journal}{Expert Systems with Applications}\/},  {\it
  \bibinfo{volume}{36}\/}, \bibinfo{pages}{5445--5449}.
\bibitem[{Zhang et~al.(2012)Zhang, Zhu, Xu \& Wan}]{zhang2012predicting}
\bibinfo{author}{Zhang, X.}, \bibinfo{author}{Zhu, J.}, \bibinfo{author}{Xu,
  S.}, \& \bibinfo{author}{Wan, Y.} (\bibinfo{year}{2012}).
\newblock \bibinfo{title}{Predicting customer churn through interpersonal
  influence}.
\newblock {\it \bibinfo{journal}{Knowledge-Based Systems}\/},  {\it
  \bibinfo{volume}{28}\/}, \bibinfo{pages}{97--104}.
\bibitem[{Zhu et~al.(2011)Zhu, Wang, Wu \& Zhu}]{zhu2011role}
\bibinfo{author}{Zhu, T.}, \bibinfo{author}{Wang, B.}, \bibinfo{author}{Wu,
  B.}, \& \bibinfo{author}{Zhu, C.} (\bibinfo{year}{2011}).
\newblock \bibinfo{title}{Role defining using behavior-based clustering in
  telecommunication network}.
\newblock {\it \bibinfo{journal}{Expert Systems with Applications}\/},  {\it
  \bibinfo{volume}{38}\/}, \bibinfo{pages}{3902--3908}.

\end{thebibliography}
}

\normalsize
\newpage
\begin{appendices}

\section{Relational Learners}\label{App:RL}
\subsection{Relational Classifiers}
Relational classifiers infer class labels for unknown nodes based on class labels of neighboring nodes and weights of links.
In particular, for each node $v_i$ in the set of unknown nodes $\mathcal{V}^U$ they estimate the probability
\[
P(l_i=c|\mathcal{N_i})
\]
for each class $c_j\in\mathcal{C}$ and some neighborhood $\mathcal{N}_i$ of $v_i$. 
In the case of customer churn the set of classes can be expressed as
\[
\mathcal{C}=\{0,1\}=:\{c_0,c_1\}
\]
and clearly
\[
P(l_i=c_1|\mathcal{N_i})=1-P(l_i=c_0|\mathcal{N_i}).
\]
Therefore, it is only necessary to calculate one of the probabilities.

\subsubsection{Weighted Vote Relational Neighbor classifier}
The weighted vote relational neighbor classifier (WVRN) considers the labels of the nodes in the first order neighborhood and determines a score for it using the weighted average, or  
\[
P(l_i=c_1| \mathcal{N}_i)=\frac{1}{Z}\sum_{v_j\in\mathcal{N}_i}w_{ij}P(l_j=c_1|\mathcal{N}_j)
\]
where $Z$ is a normalizing factor.

\subsubsection{Class Distribution Relational Neighbor Classifier}
Developed by \citet{rocchio1971relevance}, the class distribution relational neighbor classifier (CDRN) assigns a class label to a node based on the distribution of classes of neighboring nodes.  In particular, the probability for each node is defined as 
\[
P(l_i=c_1|\mathcal{N}_i)=\textrm{sim}(CV(v_i)_1,RV(c_1))
\]
where $\textrm{sim}(CV(v_i)_1,RV(c_1))$ is the normalized cosine similarity between the class vector of node $v_i$ and the reference vector of class $c_1$.  
The class vector $CV$ is defined for each node as the weighted sum of each class, or
\[
CV(v_i)_k=\sum_{v_j \in N_i} w_{ij} P(l_j=c_k), \quad k\in\{0,1\}
\]
Then, the reference vector $RV$ is the average of the class vectors of nodes with known labels, i.e.
\[
	RV(c_1)=\frac{1}{|\mathcal{V}^k_{c_{1}}|}\sum_{v_i\in \mathcal{V}_{c_1}^k}CV(v_i)_1
\] 
where $\mathcal{V}_{c_1}^k=\{v_i| v_i\in \mathcal{V}^k, l_i=c_1\}$.

Because of the particular setup for CCP where predictions are made at time $t$ for churn behaviour at $t+1$, and therefore $\mathcal{V}_{c_1}^k=\emptyset$, the reference vector can't be calculated for this time.
Instead, in a pre-training step as in \citet{verbeke2014social}, the network of a previous timeframe $t-1$ is used and compared to labels at time $t$, to create $RV$.  
As a result, when inferring the final probabilities at time $t+1$, the similarity between CV at time $t$ and the RV, created using CVs at time $t-1$, is computed.

\subsubsection{Network-Only Link-Based Classifier}
The network-only link-based classifier (NLB) was introduced by \citet{lu2003link}.
The method infers class probabilities of a node based on a logistic regression model built using the link based measures of the node as parameters, that is
\[
P(l_i=1|\mathcal{N}_i)=\frac{1}{1+e^{-\beta_0-\beta_1 CV(v_i)}}
\]
where $\beta_0$ and $\beta_1$ are the parameters from the logistic model.
\citet{lu2003link} also showed that the count link, equivalent to the normalized class vector defined as for CDRN above, is the best predictor, i.e.
\[
CV(v_i)_k=\frac{\sum_{v_j \in N_i} w_{ij} P(l_j=c_k)}{\sum_{v_j \in N_i} w_{ij}}, \quad k\in\{0,1\}.
\]
As for CDRN, pre-training using information of a previous timeframe is applied.
The coefficients $\beta_0$ and $\beta_1$ are estimated by fitting a logistic regression model to the class vectors at time $t-1$ using the known labels at time $t$.
Therefore, an additional, earlier timeframe is needed when NLB is applied.

\subsubsection{Spreading Activation Relational Classifier}
The spreading activation method was introduced by \citet{dasgupta2008social} and is frequently applied to churn prediction in telco \citep{backiel2015combining}.
It attempts to simulate the word-of-mouth effect of churn thereby spreading 'churn influence' through the network by the means of a diffusion process.
In their paper, \citet{verbeke2014social} split the spreading activation method into a relational classifier part and a collective inference method part to make it fit into their framework.

The spreading activation relational classifier (SPA RC) is similar to WVRN, in that the churn probability is defined as 
\[
P(l_i=c_1| \mathcal{N}_i)=\frac{d}{Z}\sum_{v_j\in\mathcal{N}_i} \frac{w_{ij}}{\sum_{s\in \mathcal{N}_j}w_{js}}P(l_j=c_1|\mathcal{N}_j)
\]
where the diffusion is given by the constant $d\in(0,1)$.
The difference from WVRN is that here information from the second order neighborhoods is taken into account, which gives nodes that are further away influence in the inferencing process. 

\subsection{Collective Inference Methods}
Collective inference methods are procedures which infer class labels for nodes in a network while taking into account how the inferred labels affect each other.
They define in which order the nodes are considered and how a final label or probability is determined.

Each CI applies a relational classifier, $RC$, on a combination of a network and a vector of labelled nodes. 
The network remains the same throughout the process but the labels are updated in each iteration.
At the beginning, these are the known labels at time $t$ and we denote them by $v_i^{(0)}$, where $i\in\{1,\dots,n\}$ is the number of nodes in the network.
The superscript in brackets is the number of the iteration, for which we use the index $j$.
The application of $RC$ on the label vector $v_i^{(j)}$ results in a vector of class estimates, denoted by $\hat{\textbf{c}}_i^{(j+1)}$.

\subsubsection{Gibbs Sampling}
Gibbs sampling (gib) was first described by \citet{geman1984stochastic} and was used for image recovery.  It works by applying a relational classifier to each subsequent node $v_i$ in a random ordering of $\mathcal{V}^U$. 
The result is a vector of estimated class labels $\hat{\textbf{c}}_i$ and then a value for $c_i$ is sampled from the vector such that $P(c_s=c_k|\hat{\textbf{c}}_i)=P(\hat{\textbf{c}}_i)$ and assigned to $l_i$.  
The Gibbs sampler starts by applying this method for a burn-in period with out keeping track of resulting probabilities.
Subsequently, it applies the iteration a fixed number of times, adding up the scores and averaging them at the end.

The high complexity of the original Gibbs sampler makes it slow when applied to large networks.  
Therefore, \citet{verbeke2014social} adjusted the algorithm so that the nodes were labelled concurrently instead of sequentially.  
In addition, we implemented an early stopping mechanism, as described in the paper, see Algorithm \ref{alg:gibbs}.

\begin{algorithm}
\caption{Gibbs Sampling with simultaneous labelling and early stopping}\label{alg:gibbs}
\begin{algorithmic}
\State Set $L_i \gets 0 \forall i$
\For{$j\in \{1,2,\dots,200\}$}
\State $\hat{\textbf{c}}_i^{(j+1)} \gets RC(v_i^{(j)})$
\State Sample $c_s$ from $\hat{\textbf{c}}_i$ such that $P(c_s=c_1|\hat{\textbf{c}_i})=\hat{\textbf{c}}_i(1)$
\State Set $v_i^{(j+1)}\gets c_s$
\EndFor
\For{$t\in \{1,2,\dots,2000\}$}
\State $\hat{\textbf{c}}_i^{(j+1)} \gets RC(v_i^{(j)})$
\State Sample $c_s$ from $\hat{\textbf{c}}_i$ such that $P(c_s=c_1|\hat{\textbf{c}_i})=\hat{\textbf{c}}_i(1)$
\State Set $v_i^{(j+1)}\gets c_s$
\State Set $L_i\gets L_i+c_s$
\If{ $|\sum_i L_i/j-\sum_i(L_i-c_s)/(j-1)| <$ threshold}
\State \Return $L/j$
\EndIf
\EndFor
\State \Return $L/2000$
\end{algorithmic}
\end{algorithm}

\subsubsection{Iterative Classification}
Iterative classification (IC) was in fact proposed by \citet{lu2003link} together with the link-based variables and classifier discussed above.
This  method differs from the rest, in that it assigns labels in each step of the iteration instead of scores.
Because the code has high complexity we use the modified version proposed in \citet{verbeke2014social}, as for the Gibbs sampler.  The method can be summarized in the following way: \ref{alg:ic}
\begin{algorithm}
\caption{Iterative Classification with simultaneous labelling and early stopping}\label{alg:ic}
\begin{algorithmic}
\For{$j\in \{1,2,\dots,1000\}$}
\State $\hat{\textbf{c}}_i^{(j+1)} \gets RC(v_i^{(j)})$
\State $l_i \gets c_k$ where $k=argmax(\hat{\textbf{c}}_i^{(j+1)})$
\If{ (all $l_i==0$) \textbf{or} $(\max|v_i^{(j)}-l_i| \le$ threshold$)$}
\State \Return $l_i$
\EndIf
\State $v_i^{(j+1)}\gets l_i$ 
\EndFor
\Return $v_i^{(j+1)}$
\end{algorithmic}
\end{algorithm}

\subsubsection{Relaxation Labelling}
The relaxation labelling (RL) collective inference method was proposed by \citet{chakrabarti1998enhanced} and implemented in NetKit.  
It applies a relational classifier multiple times, in each step updating the scores of the nodes, and using the new scores as class labels in the next classification iteration, see Algorithm \ref{alg:rl}.
Like in the adjusted Gibbs sampler, inferences are made simultaneously. 
\begin{algorithm}
\caption{Relaxation Labelling with early stopping}\label{alg:rl}
\begin{algorithmic}
\For{$j\in \{1,2,\dots,100\}$}
\State $\hat{\textbf{c}}_i^{(j+1)} \gets RC(v_i^{(j)})$
\State $v_i^{(j+1)} \gets \hat{\textbf{c}}_i^{(j+1)}$ 
\If{$\max|v_i^{(j+1)}-v_i^{(j)}| \le$ threshold}
\State \Return $v_i^{(j+1)}$
\EndIf
\EndFor
\Return $v_i^{(j+1)}$
\end{algorithmic}
\end{algorithm}

\subsubsection{Relaxation Labelling with Simulated Annealing}
Relaxation labelling with simulated annealing (RLSA) is a modification of RL with an additional simulated annealing with decay $\alpha$ and a constant $k\in [0,1]$, see Algorithm \ref{alg:rlsa}.
\begin{algorithm}
\caption{Relaxation Labelling with simulated annealing and early stopping}\label{alg:rlsa}
\begin{algorithmic}
\State $\beta \gets k$
\For{$j\in \{1,2,\dots,100\}$}
\State $\hat{\textbf{c}}_i^{(j+1)} \gets \beta\cdot RC(v_i^{(j)})+(1-\beta)\cdot \hat{\textbf{c}}_i^{(j)}$
\State $v_i^{(j+1)} \gets \hat{\textbf{c}}_i^{(j+1)}$ 
\If{$\max|v_i^{(j)}-v_i^{(j+1)}| \le$ threshold}
\State \Return $v_i^{(j+1)}$
\EndIf
\State $\beta\gets\beta\cdot\alpha$
\EndFor
\Return $v_i^{(j+1)}$
\end{algorithmic}
\end{algorithm}

\subsubsection{Spreading Activation Collective Inference}
In an attempt to model the word-of-mouth effect, \citet{dasgupta2008social} proposed the spreading activation method to simulate the diffusion of churn through a network.  The modification by \citet{verbeke2014social} of the CI part is the following.

The method initializes class labels as in Gibbs sampling. For each node in the set of unknown nodes, a classifier is applied, which updates the scores.  This is repeated until the aggregated change in scores does not exceed a given threshold or until a maximum number of iterations has been reached, see Algorithm \ref{alg:spaci}.
\begin{algorithm}
\caption{Spreading Activation Collective Inference Method}\label{alg:spaci}
\begin{algorithmic}
\State Set $\hat{\textbf{c}}_i^{(1)}\gets v_i^{(0)}$ \textbf{and} $j\gets 0$
\While{$j\le 100$ \textbf{and} $( \max|\hat{\textbf{c}}_i^{(j+1)}-v_i^{(j)}| \le$ threshold \textbf{or} $\sum(\hat{\textbf{c}}_i^{(j+1)}>0) > \sum(v_i^{(j)}>0)$)}
\State $\hat{\textbf{c}}_i^{(j+1)} \gets RC(v_i^{(j)})$
\State $v_i^{(j+1)}\gets \hat{\textbf{c}}_i^{(j+1)} $ 
\State $j \gets j+1$
\EndWhile
\Return $v_i^{(j)}$
\end{algorithmic}
\end{algorithm}

\section{Abbreviations}\label{app:abb}
\begin{table}[h!]
\caption{Table of Abbreviations}
\begin{tabular}{lll}
\hline
&Abbreviation&Meaning\\ \hline
\multirow{8}{*}{General}&SNA&Social Network Analysis\\
&CCP&Customer Churn Prediction\\
&CDR&Call-Detail Records\\
&RL& Relational Learner\\
&RC& Relational Classifier\\
&CI& Collective Inference Method\\
&NRC& Non Relational Classifier\\
&RFM&Recency, Frequency and Monetary \\\hline
\multirow{4}{*}{RC}&WVRN/wvrn& Weighted Vote Relational Neighbor Classifier\\
&CDRN/cdrn& Class Distribution Relational Neighbor Classifier\\
&NLB/nlb& Network Only Link Based Classifier\\
&SPA RC/spaRC/sp& Spreading Activation Relational Classifier\\\hline
\multirow{6}{*}{CI}&gib/gibbs& Gibbs Sampling\\
&IC/ic& Iterative Classification\\
&RL/rl& Relaxation Labelling\\
&RLSA/rlsa& Relaxation Labelling with Simulated Annealing\\
&SPA CI/spaCI/sp&Spreading Activation Collective Inference Method  \\
&NO&No/NO Collective Inference Method  \\ \hline
\multirow{3}{*}{NRC}&Log&Logistic Regression\\
&RF&Random Forest\\
&NN&Nerual Networks\\ \hline
\multirow{2}{*}{Measure}&MP&Maximum Profit\\
&EMP&Expected Maximum Profit\\ \hline
\end{tabular}
\end{table}
A relational learner is constructed by combining a relational classifier with a collective inference method.
This is denoted by writing CI-RC.
For example, gibbs-nlb means the combination of Gibbs sampling with the network only link based classifier.
\end{appendices}

\end{document}